# Giant gate-tunability of complex refractive index in semiconducting carbon nanotubes

Baokun Song,[†,‡] Fang Liu,[#] Haonan Wang,[§] Jinshui Miao,[†] Yueli Chen,[§] Pawan Kumar,[†,&]

Huiqin Zhang,[†] Xiwen Liu,[†] Honggang Gu,[‡] Eric A. Stach,[&] Xuelei Liang,[#] Shiyuan Liu,[‡,*]

Zahra Fakhraai,[§] Deep Jariwala[†,*]

[†]Department of Electrical and Systems Engineering, University of Pennsylvania, Philadelphia, PA 19104, USA

[‡]School of Mechanical Science and Engineering, Huazhong University of Science and Technology, Wuhan 430074, P. R. China

[#]Department of Electronics, Peking University, Beijing 100871, P. R. China

[§]Department of Chemistry, University of Pennsylvania, Philadelphia, PA 19104, USA

[&]Department of Materials Science and Engineering, University of Pennsylvania, PA 19104, USA.

ABSTRACT

Electrically-tunable optical properties in materials are desirable for many applications ranging from displays to lasing and optical communication. In most two-dimensional thin-films and other quantum confined materials, these constants have been measured accurately. However, the optical constants of single walled carbon nanotube (SWCNTs) as a function of electrostatic tuning are yet to be measured due to lack of electronic purity and spatial homogeneity over large areas. Here, we measure the basic optical constants of ultrathin high-purity (>99%)

- 1 -

semiconducting single wall carbon nanotube (s-SWCNT) films with spectroscopic ellipsometry. We extract the gate-tunable complex refractive index of s-SWCNT films and observe giant modulation of the real refractive index (~11.2% or an absolute value of > 0.2) and extinction coefficient (~11.6%) in the near-infrared (IR) region (1.3–1.55 µm) induced by the applied electric field significantly higher than all existing electro-optic semiconductors in this wavelength range. We further design a multilayer IR reflection phase modulator stack by combining s-SWCNT and monolayer $MoS_2$ heterostructures that can attain >45° reflection phase modulation at 1600 nm wavelength for < 200 nm total stack thickness. Our results highlight s-SWCNTs as a promising material system for infrared photonics and electro-optics in telecommunication applications.

INTRODUCTION

The complete knowledge of optical constants of a material is essential for its applicability in the design and development of optoelectronic, photonic, and electro-optic (EO) devices. Over the years, the knowledge of optical constants to a high degree of accuracy has been an essential component of materials development and applications research in optoelectronics. The most basic and widely known optical constants of a material arise from its electronic dispersion and are called the complex refractive index: the real part is the refractive index given by Snell's law, and the imaginary part is the extinction co-efficient. Dynamic tuning of these optical constants with external stimuli such as heat, light, strain, magnetic or electric field is the basis of optical modulation.[1–3]

The optical modulator is one of the most important components/devices in any optical interconnect link and lies at the heart of telecommunications and internet hardware. Its primary function is to modulate the attributed parameters (amplitude, phase, and polarization) of light in the time domain or frequency domain. Based on the above discussed stimuli to modulate optical constants, optical modulators can be classified as all-optical modulators, EO modulators, magneto-optic modulators, acousto-optic modulators, mechano-optical modulators, and so on.[2,3] Among them, EO modulators are the most promising due to the simplicity of applying voltages to microfabricated electrodes and ability to minimize parasitic circuit elements. The performance of EO modulators strongly depends on the EO effect of component materials.[4] This is especially important in the telecommunication region (1.3–1.5 μm) which represents the loss minima in the spectrum of optical fibers. Therefore, new semiconductor materials that demonstrate a large EO effect in this wavelength range are highly desirable.

Silicon (Si) which is the semiconductor of choice exhibits extremely weak inherent electric-refractive (ER) and electric-absorption (EA) at two important telecommunication wavelengths 1.3 μm and 1.55 μm.[1,5] This can be traced back to the natural lack of Pockels effect[5] and negligible Franz-Keldysh effect.[5] Thus, most of the silicon photonics devices rely on the plasma dispersion effect (PDE) to tune optical constants, while these silicon modulators usually need extremely high extra carrier density to achieve appropriate optical modulation.[6,7] To overcome these drawbacks of Si, Quantum Well (QW) structures of III-V semiconductors (GaAs, InP, and their alloys) have been developed that exhibit a stronger EO effect namely quantum-confined Stark effect (QCSE).[8] Modulators based on QCSE can achieve narrow-band, high-efficiency EA, however the complex fabrication requirements and CMOS compatibility still limit their widespread applications.[9] The advent of atomically-thin, van der Waals (vdW) materials, such as graphene, has created interesting possibilities for EO modulation including integration on Si and harnessing hybridization with phonon polaritons of polar materials such as SiC[10,11] and h-BN[12,13] which have proven to the superior substrates for graphene.[3,14] While the intrinsic modulation depth of monolayer (1L) graphene-based devices can only be up to ~0.1 dB due to the small absorptance 2.3%.[3] The use of waveguides or resonant optical cavities integrated with graphene can meet the performance requirements of short-range optical interconnects in Si-based integrated photonics. Beyond graphene, optical constants of two-dimensional (2D) transitional metal dichalcogenides (TMDCs) (2D-WS$_2$ and 2D-MoS$_2$) have also shown to exhibit giant tunability via application of a gate voltage due to the presence of strongly bound excitons.[15,16] However due to the visible frequency band-gaps of these materials and their excitonic transitions, they are not suitable for the telecommunication applications.

In comparison, semiconducting single wall carbon nanotubes (s-SWCNT) are a well-studied yet extremely promising material platform that can be used to design infrared (IR) EO devices. Carbon-nanotubes are structurally and electronically quantum-confined in one dimension, and therefore possess intrinsically discrete absorption singularities,[17] large exciton binding energy,[18] gate-tunable optical properties,[19,20] as well as good CMOS compatibility,[21] which are superior to their 2D counterparts in many respects. Electro-absorption properties of s-SWCNT have been studied previously,[21,22] however, to our best knowledge, a reliable and systematic study on the gate-tunable inherent optical constants of s-SWCNTs, such as refractive index and extinction coefficient, has not yet been reported. This is mainly because of the lack of high electronic purity in the samples and the difficulty in controlling the film thickness and uniformity over a large area. The lack of electronic purity introduces high conductance and screening from the metallic nanotube impurities which prevents reliable gate modulation.[22] However, over the past decade several advances have been made in purification and deposition of s-SWCNT films which renders this possible in our study.[23–29]

In this paper, we determine the gate-tunability of the basic optical constants of ultrathin s-SWCNT films, including refractive index $n$, extinction coefficient $\kappa$, and absorption coefficient $\alpha$. We find that the optical constants of s-SWCNT in near-IR region show striking changes and excellent sensitivity with the gate voltage ($V_G$). The electric field-induced changes in $n$ and $\kappa$ ($\Delta n$, $\Delta \kappa$) of s-SWCNT is much greater than those of traditional semiconductors, such as Si, Ge, and GaAs. The remarkable $\Delta n$, $\Delta \kappa$, excellent Si-CMOS compatibility,[27,30–32] and low fabrication complexity renders s-SWCNTs as attractive candidates for designing IR photonic devices specially in the telecommunications band. Beyond determining and comparing field and density tunable optical constants, we show an ultrathin IR gate-tunable reflection phase modulator stack with a superlattice structure of (1L-MoS$_2$/s-SWCNT)$_5$/MoS$_2$/Au, which shows theoretical phase change exceeding 45° ($\pi$/4) for ~ 100 nm active layer s-SWCNT phase modulator in the stack for $\lambda$ = ~1600 nm. This phase tuning was achieved without the need for any nanostructuring or nanopatterning to a metasurface which is noteworthy. The gate-tunable optical constants of s-SWCNTs reported here could provide theoretical guidance for next-generation Si CMOS-compatible photonics and optoelectronics devices as well as IR photodetectors, sensors, modulators, and energy harvesters.

RESULTS AND DISCUSSION

We deposit ultrathin films of s-SWCNT directly on a 4-inch SiO$_2$/Si by dip-coating using a chloroform-dispersed high-purity (>99%) semiconducting SWCNT dispersion (prepared by Arc discharge method).[23,24] The optical photograph and scanning electron microscopy (SEM) image of the s-SWCNT film specimen are shown in **Figure 1(a,b)**, where the uniform optical contrast and flat surface morphology suggest that the film is macroscopically and microscopically uniform. The dark lines that appear on the film surface in SEM micrograph (Figure 1(b)) are attributed to a small number of bundles. Figure 1(c) shows a representative atomic force microscopy (AFM) topography image of the film which indicates a random structure of the network and root-mean-square surface roughness ($S_q$) of the film is estimated to be ~1.78±0.51 nm. These roughness values are much smaller than the spot size of our measurements (~2 mm) and wavelength of light (371–1687 nm). The taller and wider tube-like features in the AFM image can be interpreted as overlapping two or more tubes and bundles (thick white lines and nodes in Figure 1(c)). The nominal diameters of the suspended s-SWCNTs in the solution are about 1.4 nm–1.6 nm, and the average length is ~1.5 μm –2 μm based on an earlier published report.[23] We have further characterized the s-SWCNT films using Raman spectroscopy, which is sensitive to electronic character and chirality. A representative Raman spectrum of the SWCNT film is shown in Figure 1(d). Six typical Raman modes/features can be identified: radial breathing mode (RBM) at ~153 cm$^{-1}$, bundle breathing-like mode (BBLM) at ~ 420 cm$^{-1}$,[32] D-band at ~1328 cm$^{-1}$, weak E$_2$ feature at ~1561 cm$^{-1}$ associated with the symmetry phonons,[33,34] G$^+$ feature at ~1591 cm$^{-1}$ resulting from the in-plane vibrations along the tube axis (longitudinal optical (LO) phonon mode), and a diameter-dependent G$^-$ feature at ~1574 cm$^{-1}$ for in-plane vibrations along the circumferential direction (transverse optical (TO) phonon mode).[33] The G-band with clear Lorentzian line-shape (Figure S1(f)) indicates that our SWCNT film comprises of only semiconducting tubes.[33] The appearance of weak Si mode at 303 cm$^{-1}$ suggests that the Raman responses were predominantly from the s-SWCNTs.[33] Additional Raman analysis suggests that the film predominantly comprises of two types of nanotubes of diameters $d_{t1}$ and $d_{t2}$ at about 1.62 nm and 1.55 nm corresponding to (17,6) and (16,6) chirality respectively. Both of these chiral indices belong to semiconducting tubes in the families of 40 and 38.[35] Additional analysis and calculations are provided in Supporting Information, Figure S1 and S2. Finally, we also measure the absorbance spectrum of the nanotube solution in chloroform used for deposition of the film (Figure 1e). The absorbance spectrum shows three prominent semiconducting absorption bands ($S_{22}$, $S_{33}$, and the rising side to the right of the $S_{11}$) as labeled in Figure 1(e).[25,36,37] There is no obvious or discernable metallic transition peaks $M_{11}$ (~615 nm) in the absorbance spectrum,[36] further suggesting that the SWCNT film mainly comprises of semiconductor tubes. Further information about the sample fabrication process and quality characterization can be found in Ref. 23 and Supporting Information.

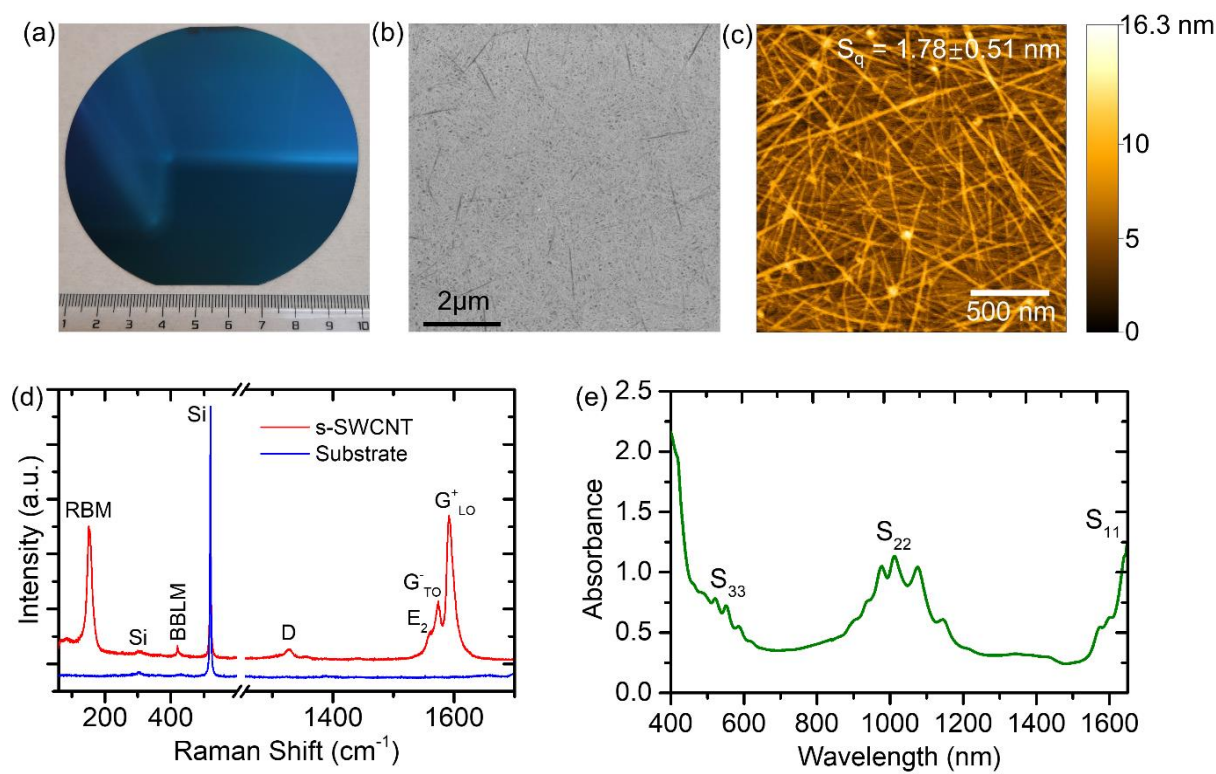

**Figure 1.** (a) Optical photograph of a 4-inch s-SWCNT coated thin-film on $SiO_2$/Si wafer. (b,c) Secondary electron scanning electron micrograph and topographic image acquired using atomic force microscopy of the s-SWCNT film, the scanning area is 2 μm × 2 μm. (d) Typical Raman spectrum of the film on $SiO_2$ (324 nm)/Si wafer. (e) Absorbance spectrum of the s-SWCNT dispersion in chloroform.

Upon structural and basic optical characterization of the s-SWCNT films, we determine the gate-tunable complex refractive index via spectroscopic ellipsometry (SE) (Figure 2). The inset in Figure 2(a) illustrates the measurement configuration/schematic of the experiment, where a gold electrode (L×W×H: 0.5 mm×0.5 mm×50 nm) is deposited onto the surface of s-SWCNT film via thermal evaporation through a stencil shadow mask. The ellipsometry spot is positioned away from this gold electrode to completely lie in the s-SWCNT film region while the Au electrode is bonded. The degenerately $n$-type doped Si underneath the 324-nm-thick SiO$_2$ is used as the back gate. When V$_G$ changes from −50 V to 0 (0 to +50 V), extra holes (electrons) will be injected into the s-SWCNT film and modulate the optical properties. Subtle changes in the interaction between light and s-SWCNT film induced by carrier injection can be collected by the ellipsometer and recorded in a set of multiangle ellipsometric spectra [$\Psi(\lambda)$, $\Delta(\lambda)$] (solid balls in Figure 2(a,b)). The measured ellipsometric spectra are fitted using theoretical plots (lines in Figure 2(a,b)) calculated from the transfer matrix method (TMM).[38] The fitting goodness is evaluated by the root-mean-square-error (RMSE). It is calculated by

$$RMSE = \sqrt{\frac{1}{2p-q}\sum_{i}^{p}[(\Psi_{exp}^{i} - \Psi_{cal}^{i})^2 + (\Delta_{exp}^{i} - \Delta_{cal}^{i})^2]} \times 1000 \,, \text{ where, } p \text{ and } q \text{ refer to the number}$$

of wavelengths and the number of fitting parameters, and the subscripts 'exp' and 'cal' denote the experimental and calculated data. Regardless of whether the s-SWCNT film is in a finite bias or floating or grounded, the value of RMSE in the ellipsometric analysis is as low as ~4.83, meaning that the analysis is reliable. For clarity, we only show the fitting results of s-SWCNT for electrically floating (unbiased) films (V$_G$ = 0). The ellipsometric spectra and related analysis results of s-SWCNT under different $V_G$ can be found in Figure S4 and S5 of Supporting Information. The thicknesses of s-SWCNT and $SiO_2$ determined by SE are 2.04 nm and 324 nm, which is consistent with the extracted results from the Raman spectrum. Further principles, measurement processes, and data analyses about SE measurement are available in Supporting Information and previous publications.[39–43]

Figures 2(c) and 2(d) demonstrates the gate-dependent complex refractive index of s-SWCNT films. The three semiconducting absorption peaks ($S_{11}$–$S_{33}$) can be observed again in the extinction coefficient spectrum, and the wavelength positions of $S_{11}$–$S_{33}$ are consistent with the results of absorbance spectrum of the dispersion in chloroform (Figure 1(e)). When $V_G$ changes from −50 V to +50 V, $n$ shows a significant decrease, on the contrary, $\kappa$ exhibits a slight increase in the near-IR spectral region. This observation is in agreement with the evolution rules of $\Psi$ and $\Delta$ spectra as shown in Figure S4(c) and S4(d). $\Delta$ spectrum is more sensitive to $V_G$ than $\Psi$ spectrum ($\delta\Delta = 3°$ and $\delta\Psi = 0.6°$ when $V_G$ changes from −50 V to +50 V), which leads to the significant change of $n$. The gating tunability of $n$ and $\kappa$ can be interpreted by PDE.[5,22] Semiconducting SWCNTs are nominally p-doped in ambient which means they have an excess of holes.[44] The positive (negative) $V_G$ decreases (increases) the hole concentration in the s-SWCNT film. Therefore, at high positive $V_G$ the s-SWCNT film is nearly intrinsic with minimal excess free carriers. This minimizes the Pauli blocking[45] and enhances the oscillator strength of $S_{11}$ increasing the $\kappa$.[21] Conversely, a negative $V_G$ heavily p-dopes the sample and reduces the strength of the $S_{11}$ transition.[21,22] We apply the above relationship to explain the $V_G$-modulated intensity change of $\kappa$ (Figure 2(d)). When $V_G$ changes from –50 V to +50 V, the density of holes will gradually reduce to intrinsic levels. This also corroborates with p-type character of thin-film transistors made from similar samples by the authors in the past.[23,24] Thus, $\kappa$ increases monotonously as $V_G$ changes from –50 V to + 50 V Prior studies have also observed similar changes in the absorbance spectra of CNTs.[21,22] The refractive index $n$ of s-SWCNT gradually decreases with the increasing $V_G$. This is in agreement with the relation between the injected carrier concentration and the change in $n$ ($\Delta n$),[5] confirming that the evolution trend of $n$ can also be well explained by PDE. A slight gate-dependent redshift appears in the IR portion of $n$ spectra at ~1600 nm (close to $S_{11}$ transition) as indicated by a blue arrow in Figure 2(c). We infer that this shift is associated with the Stark effect in s-SWCNT, which can trigger some additional absorption below $S_{11}$ and $S_{22}$ transitions.[20]

Here, it is worth mentioning that that complex refractive index of s-SWCNTs depends on the diameter of the nanotubes, i.e., ER and RA should be different for the s-SWCNT films with different diameters. In order to verify this, we also prepared a high-purity HiPCO grown s-SWCNTs film on $SiO_2$/Si substrate as shown in Figure S6, whose diameter is in the range of ~0.8 nm–1.1 nm, with an average length of about 1.5 μm. The complex refractive index of HiPCO s-SWCNT films is determined by SE and shown in Figure S6(b,c). As compared with the Arc discharge s-SWCNTs, the complex refractive indices, optical transition energies and optical bandgaps of HiPCO s-SWCNT films are larger, which is expected due to the stronger 1D quantum confinement effect and tighter diameter distribution in smaller diameter tubes.[46,47]

In addition to the diameter, the optical constants may also vary with the length of the s-SWCNTs. A larger average length of the s-SWCNT could increase the mean free path and mobility of the axially moving carriers, ultimately influence the strength of optical constants, such as optical conductivity and absorption.[48]

Figures 2(e) and 2(f) display the relative percentage changes of $n$ and $\kappa$ in the near-IR region with the external applied electric field estimated via parallel place capacitive coupling (see below). As compared with the extremely weak electric-field and carrier density induced changes of $n$ and $\kappa$ in traditional semiconductors,[1,49] the relative changes of $n$ and $\kappa$ in s-SWCNT can reach striking ~15% or an absolute value of > 0.2. More importantly, the gate-tunability of $n$ and $\kappa$ are remarkably high at the two most useful telecommunication wavelengths i.e. 1.3 μm and 1.55 μm.

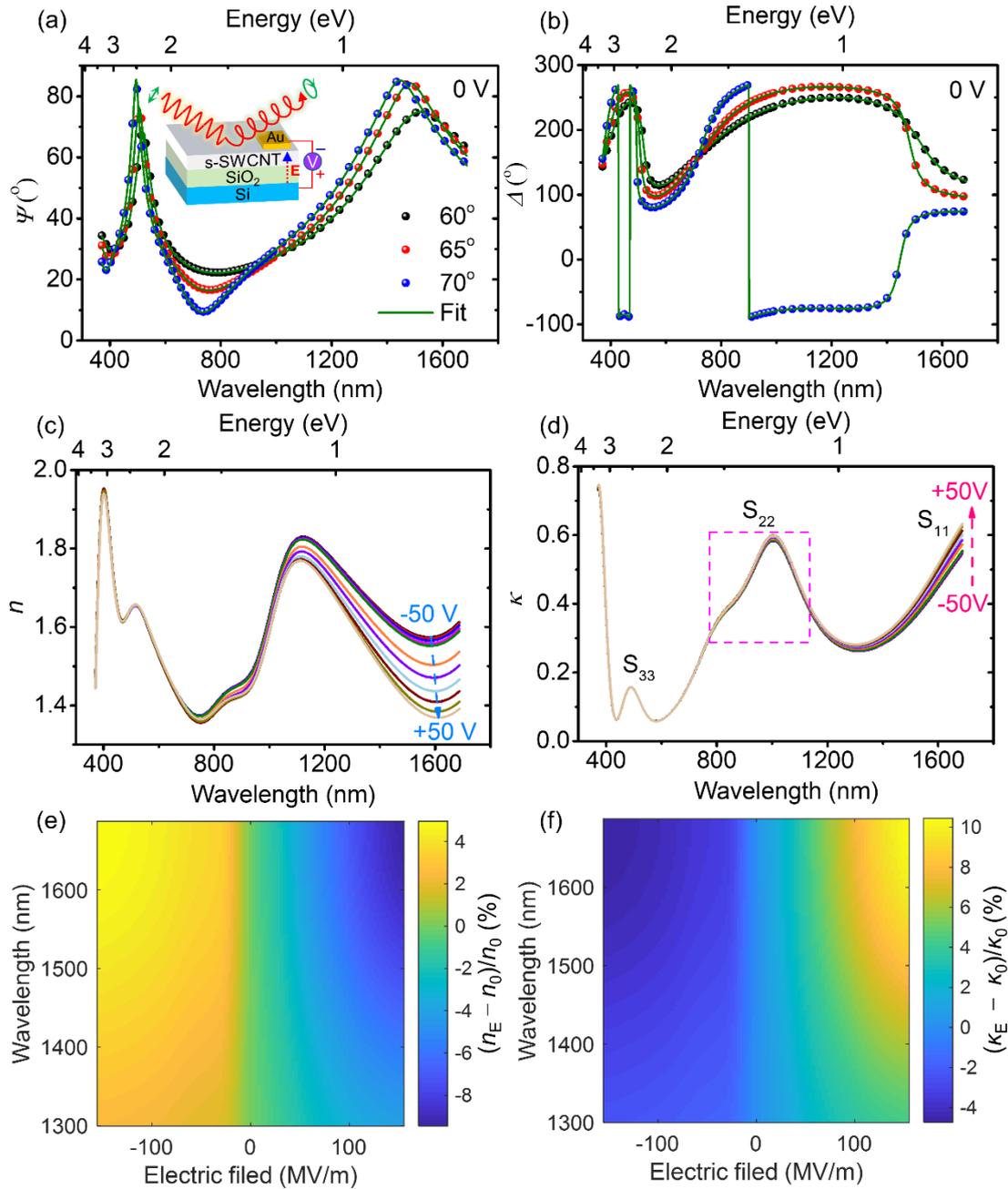

**Figure 2.** (a,b) Measured (spheres) and best-fits (lines) of multiangle incident (60°, 65°, and 70°, w.r.t normal incidence) ellipsometric spectra of s-SWCNT film, for clarity, only fitting results for zero applied volts are plotted. The inset in (a) shows the schematic of SE measurement that is used to measure the s-SWCNT specimen under applied gate voltage. (c) Gate-tunable $n$ of s-SWCNT, the voltage changes from −50 V to +50 V with a step size of 10 V. The large modulation in $n$ between 1100-1687 nm is apparent from the plots. (d) Gate-tunable $\kappa$ of same voltage and spectral range with relatively small modulation between 1200-1687 nm. The pink dash square in (d) indicates the shoulder and peak region that is attributed to the $S_{22}$ transition. (e,f) Relative changes of $n$ and $\kappa$ with the applied electric field over the near-IR region (1300 nm–1687 nm). $n_E$ ($\kappa_E$) and $n_0$ ($\kappa_0$) stand for the refractive indices

(extinction coefficients) in presence and absence of an applied electric field respectively.

Figures 3(a) and 3(b) illustrates the evolution of $\Delta n$ and $\Delta \alpha$ (where $\alpha = 4\pi\kappa/\lambda$ is the absorption coefficient in cm$^{-1}$) of s-SWCNT at 1.3 μm and 1.55 μm with the carrier density $N_c$ induced by $V_G$. $N_c$ is calculated by $N_c = \varepsilon_0 \varepsilon_{SiO_2} V_G / d_{SiO_2}$, where $\varepsilon_0$ and $\varepsilon_{SiO_2} = 3.9$ denote the free-space permittivity and static dielectric constant of SiO$_2$,[50] $d_{SiO_2} = 324$ nm. The trends of $\Delta n$ and $\Delta \alpha$ as a function of $N_c$ are opposite. The overall changes at 1.55 μm are greater (0.2 for $\Delta n$ and 4500 cm$^{-1}$ for $\Delta \alpha$), as compared to those of 1300 nm, which can be mainly attributed to the greater influence from the enhanced S$_{11}$ transition. Further, s-SWCNTs are also known to exhibit gate-tunable excitons and trions in the near-IR region which may also influence this tunability.[51]

Figures 3(c) and 3(d) shows comparative values of $\Delta n$ and $\Delta \alpha$ for our s-SWCNT films with some traditional semiconductors, including Si,[7,52] Ge,[53] GaAs,[53] and InP,[49] as well as InGaAsN single QW[8] and Ge/SiGe multiple QW structures[54] which have been studied or used for EO modulators. We choose to selectively display some of the largest $\Delta n$ and $\Delta \alpha$ reported by prior literatures to make a fair comparison.[7,8,49,52–54] It is obvious from the plots that $\Delta n$ and $\Delta \alpha$ of s-SWCNTs are much more sensitive to the applied electric field than those of conventional semiconductors and EO material systems. With the increase of electric field, $\Delta n$ and $\Delta \alpha$ of s-SWCNT, Ge, and GaAs decrease uniformly. Comparatively speaking, the magnitude of of $\Delta n$ and $\Delta \alpha$ of Si is very small (~$10^{-6}$–$10^{-5}$ and ~0.05–5 cm$^{-1}$),[7] which is associated with the inherent lack of Pockels effect, and the weak Franz–Keldysh effect.[5] InP is another typical material platform for integrated photonics.[55] As shown in Figure 3(c), the intensity of $\Delta n$ of InP ($\sim 5 \times 10^{-5}$) is a little bit larger than that of Si. By exploiting the quantum confinement induced concentration of density of states, QCSE has also been used as the EO effect to design quaternary InGaAsN QWs based modulators,[8] however, their compatibility with the Si CMOS is far from ideal.[9] Likewise, Ge/SiGe QWs have also been used in EO modulators,[9,54] which help overcome the challenge of Si CMOS integration while exhibiting relatively high $\Delta n$ ($\sim 10^{-4}$)[54] and $\Delta\alpha$ ($\sim 2 \times 10^{3}$ cm$^{-1}$) when the applied voltage changes from 0 to 4 V).[9] The tunability of s-SWCNT refractive index is far superior even compared to ferroelectric electro-optic materials such as Lithium niobite (LiNbO$_3$). LiNbO$_3$ is a classical material platform that have been widely adopted in phase modulators in telecommunications due to the highly efficient linear EO effect and good temperature stability.[56] Still, the index change ($\Delta n_e$) of LiNbO$_3$ can only reach $\sim 0.02$ (calculated by $\Delta n_e = -0.5 n_e^3 r_{33} \mathbf{E}$, $r_{33}$ means the element of electro-optic tensor, and $n_e$ is from Ref. 57) at 633 nm when a 100 MV/m electric field is applied along the crystal symmetry axis (z-axis).[58] Likewise, photorefractive polymers with low glass transition temperatures also exhibit high refractive index modulation ($\sim 0.007$ at 90 MV/m)[59] again, much lesser compared to s-SWCNTs. Recently, some studies have also observed ER and EA effects in semiconducting 2D TMDCs[15] and black phosphorus (BP)[45]. While, the ER and EA effects in semiconducting 2D TMDCs are significant, the effects only appear at or around the excitonic transition peaks which lie in the visible part of the spectrum. This limits the EO effect and its applicability in the infrared region for semiconducting 2D

TMDCs. Likewise, a strong and anisotropic EO effect has also been reported for ultrathin BP in our previous work. While the magnitude of effect is large, the innate chemical instability of BP under ambient conditions combined with lack of large area, cm$^2$-scale samples hinders its applicability to some extent in addition to preventing extraction of accurate optical constants.

In order to make a fair comparison on the same scale, we have normalized the $\Delta n$ and $\Delta \alpha$ values over the positive electric field region to % change values in Figure 3(e) and 3(f), respectively. The large % variation of s-SWCNTs in comparison to all others is evident. The large % change in $\alpha$ of silicon can be mainly attributed to the near-zero $\alpha_0$ in the wavelength of 1070 nm. Furthermore, we also compare the $\Delta n$ and $\Delta \alpha$ of s-SWCNT with those of monolayer graphene (Figure 3(e,f)).[60] It is found that the $\Delta n$ of s-SWCNT thin-film with applied electric field in the two telecommunication bands is comparable with that of monolayer graphene, and with a clear monotonous trend, which is conducive to the design, development and operation of the photonics devices. In summary, as compared to most of the traditional semiconductors and artificial QW structures, the mass production of s-SWCNT is relatively easy and in recent years the purification and deposition has also been perfected. Further, their $\Delta n$ and $\Delta \alpha$ are far superior especially in the telecommunication bands. Further, they possess good thermal stability, are fully compatible with Si CMOS based hetero-integration and are hence promising candidates for integrated-photonics and optoelectronics.[22,61] The specific values of $\Delta n$ and $\Delta \alpha$ of these material systems shown in Figure 3(c,d) have been summarized in Table S2.

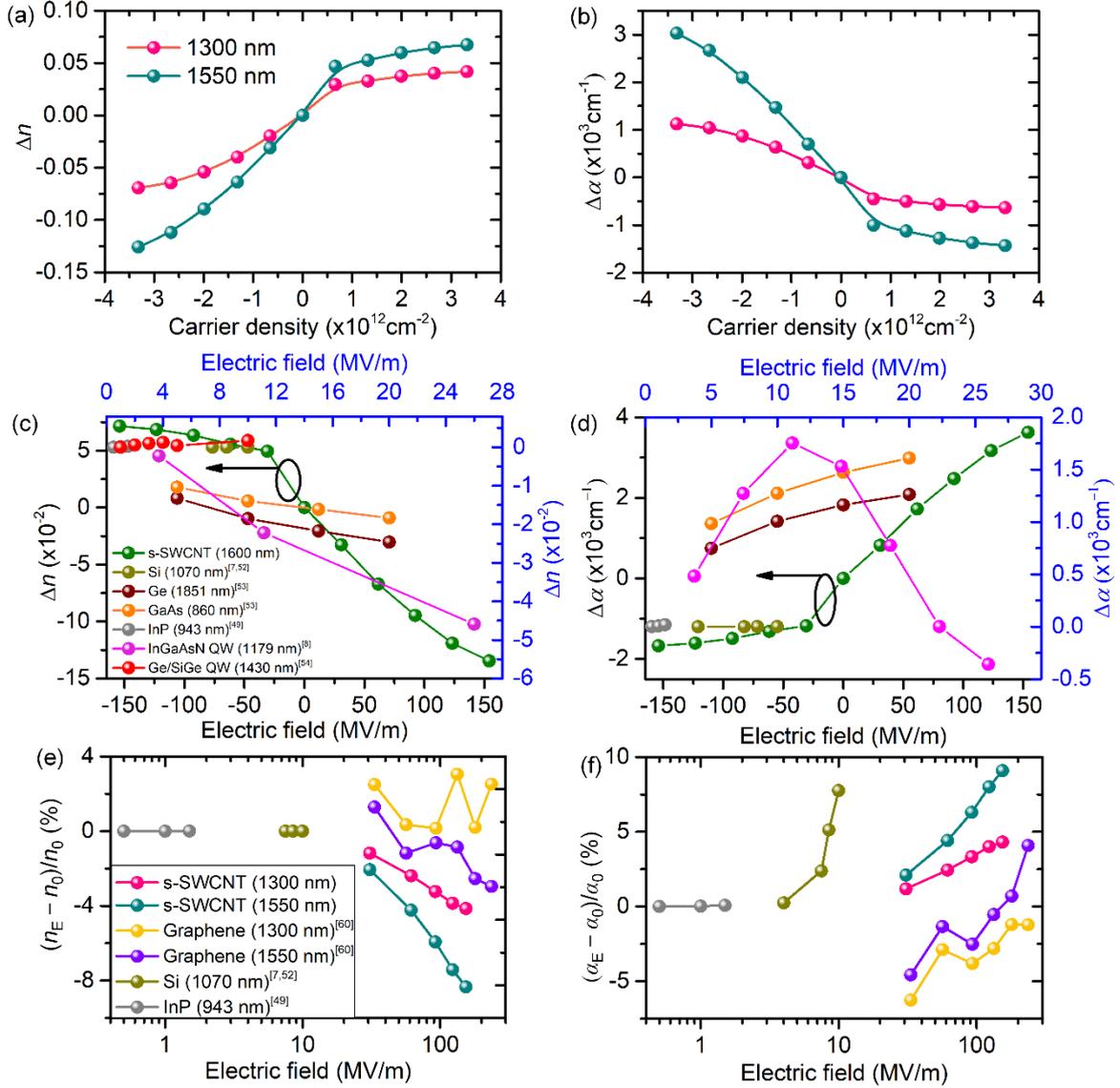

**Figure 3.** (a,b) Carrier concentration-dependent $\Delta n$ and $\Delta \alpha$ of s-SWCNT at two critical telecommunication wavelengths 1.3 µm and 1.55 µm, the positive '+' (negative '−') sign in $x$-axes indicates the injection of holes (electrons) into the s-SWCNT film. (c,d) $\Delta n$ and $\Delta \alpha$ of s-SWCNT films, some traditional semiconductor materials and QW structures used in EO modulators. Only the largest $\Delta n$ and $\Delta \alpha$ reported by prior literatures are shown to make a fair comparison Further, the $\Delta n$ and $\Delta \alpha$ of s-SWCNT films are the largest at 1600 nm in the measured spectral region. The range of electric field over which the s-SWCNT films were measured is larger than other materials hence plotted on the bottom x-axis and left y-axis for fair comparison. The (wavelengths) in figure legendindicates the spectral position from where the $\Delta n$ and $\Delta \alpha$ are obtained. (e,f) Relative changes of $n$ and $\alpha$ with the applied electric field of s-SWCNT, graphene, silicon and InP respectively. $n_E$ ($\alpha_E$) and $n_0$ ($\alpha_0$) stand for the refractive indices (absorption coefficients) in presence and absence of an applied electric field respectively.

Finally, in order to intuitively illustrate the potential applications of the ultrathin s-SWCNT films and their unique optical constants, we propose and design an EO reflection phase modulator in the IR spectrum based on the s-SWCNT and MoS$_2$ as alternating layers in a superlattice structure [(1L-MoS$_2$/s-SWCNT)$_N$/MoS$_2$/Au], where $N$ refers to the number of the unit-cell (1L-MoS$_2$/s-SWCNT). As shown in Figure 4(a), bottom MoS$_2$ film, s-SWCNT film, and gold substrate are combined to form a structure analogous to a resonant Fabry-Perot cavity. In the near-IR region, the few-nm thick MoS$_2$ can be regarded as a lossless dielectric ($\kappa \approx 0$).[62] MoS$_2$ is chosen for its vdW nature, high sub-gap index and literature precedent in combining it with s-SWCNT films in vdW heterostructure devices.[44] According to the electromagnetic wave interference theory,[63] a MoS$_2$ film with the minimum thickness of around $\lambda/4n_{\mathrm{MoS_2}}$ ($n_{\mathrm{MoS_2}}$ is the refractive index of bulk MoS$_2$) together with the perfect reflector Au substrate will form an simple, lossless Fabry-Perot cavity in the near-IR part of the spectrum.[64] In this case, the normal incident light reflected by the Au/MoS$_2$ interface will consequently produce a phase shift at the air/s-SWCNT interface after being modulated by the MoS$_2$ film, thereby laying the foundation of designing a sensitive EO phase modulator[65] without the need of nanostructuring or creating a metasurface. We perform a simple parameter sweep in a thin-film transfer matrix simulation and determine that when the thickness of bottom MoS$_2$ film ($t_1$) is ~72 nm, a maximum average reflection phase change −2.37° can be achieved in the s-SWCNT (2.04 nm) lying atop the MoS$_2$. The average reflection phase change is defined as $[\sum_{i=1}^{M}(\theta_i^{+50V} - \theta_i^{-50V})]/M$ , where $\theta_i^{+50V}$ and $\theta_i^{-50V}$ mean the reflection phases at the $i$th wavelength when the applied voltages are +50 V (+154 MV/m) and –50 V (–154 MV/m), $M$ refers to the number of wavelengths (See Figure S7(a) in Supporting information). In this design the overall maximum average phase change will increase proportionately when we increase $N$ (See Figure S7(a–c) in Supporting information). Furthermore, the influence of the layer number ($L$) of MoS$_2$ in the unit-cell on the gate-tunable phase change is also considered in the simulation (Figure S7(d)). The optical constants and thickness (0.6 nm) of 1L-MoS$_2$ used here are taken from Ref. 41.

The performance of the superlattice stack IR reflection phase modulator can be further improved by increasing the thickness of s-SWCNT (Figure 4(b–d)). Here, we assume that the in-plane complex refractive indices $N_o$ of s-SWCNT with different thicknesses are approximately equal to the isotropic complex refractive index shown in Figure 2(c,d) (Detailed discussion can be found in Supporting Information). As illustrated in Figure 4(c,d), when the thickness of s-SWCNT increases to 20 nm, the maximum reflection phase change of the multilayer stacking phase modulator can be over –45°. This $\pi/4$ phase shift in a simple multilayer stack without the need for creating a nanostructured metasurface is remarkable. Further, given that the total active layer s-SWCNT thickness is ~ 100 nm for our optimized stack of $N = 5$, the above phase modulation at $\lambda = 1600$ nm (~ 16 × the active medium thickness) is further noteworthy. Further, we have also designed an infrared absorber with maximum absorption approaching ~0.9 at 1550 nm based on the same structure ($N = 5$) shown in Figure 4(a). See Figure S8 for more details. This absorption is again significant for IR light harvesting. Further the tunable absorption properties of the s-SWCNTs reported here are expected to serve as new design inspirations for saturable absorbers in IR lasers.[66]

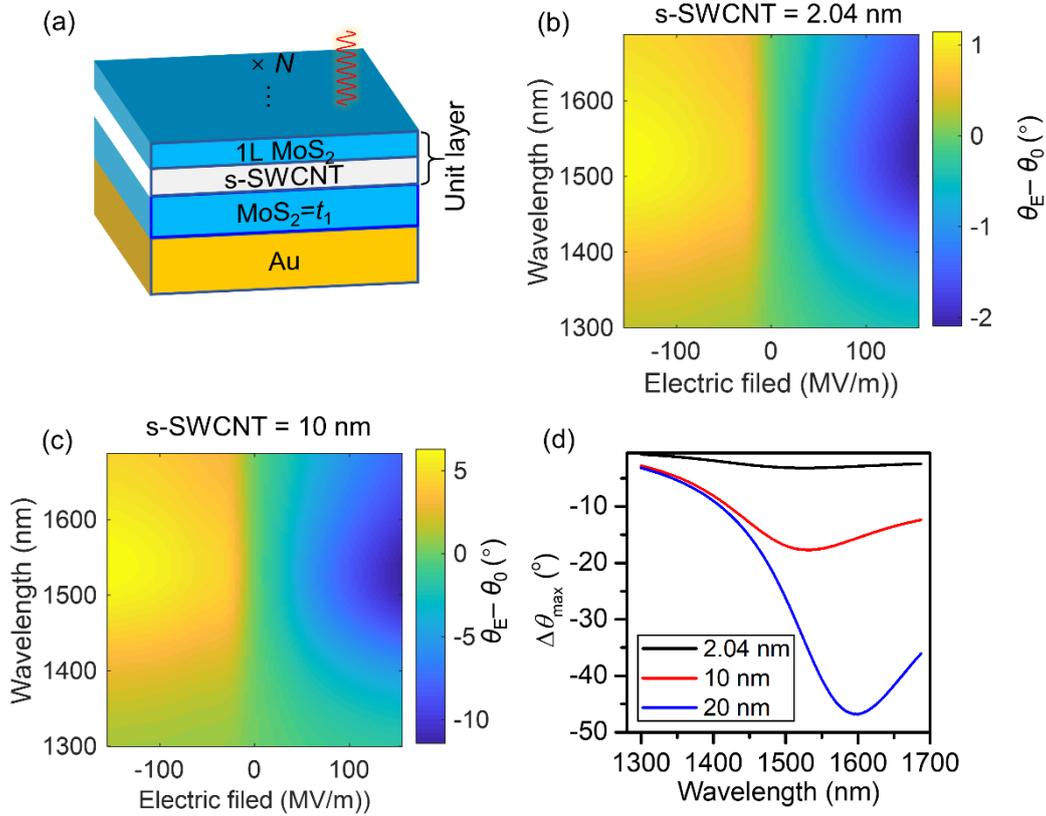

**Figure 4.** (a) Schematic of IR reflection phase modulator based on the s-SWCNT, MoS$_2$, and Au substrate. $N$ and $t_1$ represent the number of unit cell and the thickness of bottom MoS$_2$ film. (b,c) Simulated gate-tunable reflection phase changes of structure (a) with 2.04 nm and 10 nm s-SWCNT, $\theta_E$ and $\theta_0$ mean the reflection phases when the applied electric fields are $E$ and 0. (d) Simulated maximum reflection phase differences as the applied electric filed changes from −154 MV/m to +154 MV/m, where the thicknesses of s-SWCNT thin-films in the unit cell are 2.04 nm, 10 nm, and 20 nm, respectively. The $N$ used in the simulation processes of (b–d) is 5.

## CONCLUSIONS

In summary, the gate-tunable complex refractive index of the ultrathin high-purity s-SWCNT films is investigated by SE. We observe that the complex refractive index of the s-SWCNT films shows a variation of up to 15% in the near-IR region as V$_G$ changes from −50 V to +50 V. This large change can be mainly attributed to the intense PDE and exciton effect. We further find that the s-SWCNT films shows the highest sensitivity to change in optical constants ($\Delta n$ and $\Delta \alpha$) as a function the externally applied electric fields, at the telecom wavelengths of 1.3 μm and 1.55 μm in comparison to all bulk and conventional EO semiconductors and insulators. Finally, a multilayer superlattice stack-based IR reflection phase modulator based on s-SWCNT and 1L-MoS$_2$ is designed which achieves a >45° EO reflection phase modulation in the near IR. Our results suggest that high purity s-SWCNTs are very promising as highly tunable EO materials particularly in the telecommunications range. Given the magnitude of index modulation observed, our results should promote the development of SWCNT-based IR tunable optoelectronics devices, such as modulators, limiters, and saturable absorbers.

METHODS

**Preparation of s-SWCNT:** The ultrathin s-SWCNT films are deposited onto a 4-inch SiO$_2$/Si by dip-coating using a chloroform-dispersed high-purity (>99%) semiconducting SWCNT dispersion detailed in our prior publications.[23,24]

**Characterization of s-SWCNT:** The typical Raman spectrum of s-SWCNT films was measured by an integrated confocal spectrum testing and analysis system. The scattered Raman signals were collected by a microscope objective (Olympus SLMPLN 100×) and analyzed using a grating spectrometer coupled to a Si focal plane array (FPA) detector. All these instruments are integrated into the LabRAM HR Evolution confocal Microscope. The wavelength of the excitation laser is 633 nm (1.96 eV) and the lateral size of the measurement spot is less than 1 μm. The spectroscopic ellipsometer used in our experiment is M-2000 type spectroscopic ellipsometer purchased from J.A. Woollam Company, whose detector spectral range is 371 nm–1687 nm (0.73 eV –3.34 eV). The multi-incidence measurement mode (angle of incidence (AOI): 60°, 65°, and 70°) was adopted to measure the s-SWCNT films.

ASSOCIATED CONTENT

**Supporting Information**

The Supporting Information is available free of charge at https://pubs.acs.org/doi/xx.xxxx/acsphotonics.xxxxxxx. Raman analysis of s-SWCNT film; Spectroscopic Ellipsometry of s-SWCNT specimen; Complex refractive indices of Arc discharge and HiPco s-SWCNT films; Gate-tunable ellipsometry spectra of s-SWCNT film; Optimization results of (1L-MoS$_2$/SWCNT)$_5$/MoS$_2$/Au near-IR gate-tunable reflection phase modulator; Optimization results of (1L-MoS$_2$/SWCNT)$_5$/MoS$_2$/Au near-IR absorber; Tabulated $\Delta n$ and $\Delta \kappa$ of s-SWCNT, some traditional semiconductors, and QW structures. (PDF).

AUTHOR INFORMATION

**Corresponding Author**

*E-mail: shyliu@hust.edu.cn
*E-mail: dmj@seas.upenn.edu

**ORCID**
Baokun song: 0000-0002-8184-5616
Fang Liu:
Haonan Wang: 0000-0003-2047-5380
Jinshui Miao: 0000-0002-7571-2454
Yueli Chen:

Pawan Kumar:
Huiqin Zhang:
Xiwen Liu:
Honggang Gu: 0000-0001-8812-1621
Eric A. Stach: 0000-0002-3366-2153
Xuelei Liang:
Shiyuan Liu: 0000-0002-0756-1439
Zahra Fakhraai: 0000-0002-0597-9882
Deep M. Jariwala: 0000-0002-3570-8768

**Notes**

The authors declare no competing financial interest.

ACKNOWLEDGEMENTS

D.J. acknowledges primary support for this work by the U.S. Army Research Office under contract number W911NF-19-1-0109. D.J. and E.A.S also acknowledges support from National Science Foundation (DMR-1905853). D.J., E.A.S and Z.F. acknowledge partial support from University of Pennsylvania Materials Research Science and Engineering Center (MRSEC) (DMR-1720530). Huiqin Zhang acknowledges support from the Vagelos Institute of Energy Science and Technology graduate fellowship. Baokun Song acknowledges the support from China scholarship council (CSC). Baokun Song, Honggang Gu and Shiyuan Liu acknowledge the National Natural Science Foundation of China (51525502 and 51775217). A portion of this work was carried out at the Singh Center for Nanotechnology at the University of Pennsylvania which is supported by the National Science Foundation (NSF) National Nanotechnology Coordinated Infrastructure Program grant NNCI-1542153.

**For Table of Contents Use Only**

**Title:** Giant gate-tunability of complex refractive index in semiconducting carbon nanotubes

**Authors:** Baokun Song, Fang Liu, Haonan Wang, Jinshui Miao, Yueli Chen, Pawan Kumar, Huiqin Zhang, Xiwen Liu, Honggang Gu, Eric A. Stach, Xuelei Liang, Shiyuan Liu, Zahra Fakhraai, Deep Jariwala

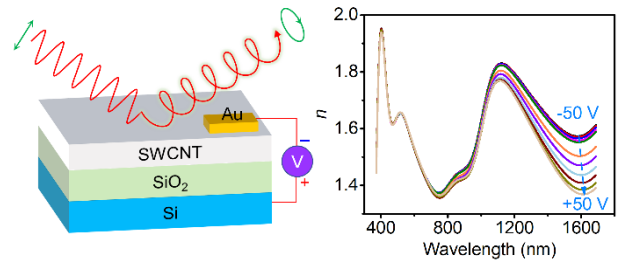

**TOC Figure:** The gate-tunable complex refractive index of the ultrathin high-purity (>99%) s-SWCNT films is investigated by spectroscopic ellipsometry. The complex refractive index of the s-SWCNT films shows a remarkable variation in the near-IR region as $V_G$ changes from −50 V to +50 V.

**Supporting Information**

**Giant gate-tunability of complex refractive index in semiconducting carbon nanotubes**

Baokun Song,[†,‡] Fang Liu,[#] Haonan Wang,[§] Jinshui Miao,[†] Yueli Chen,[§] Pawan Kumar,[†,&]

Huiqin Zhang,[†] Xiwen Liu,[†] Honggang Gu,[‡] Eric A. Stach,[&] Xuelei Liang,[#] Shiyuan Liu,[‡,*]

Zahra Fakhraai,[§] Deep Jariwala[†,*]

[†]Department of Electrical and Systems Engineering, University of Pennsylvania, Philadelphia,

PA 19104, USA

[‡]School of Mechanical Science and Engineering, Huazhong University of Science and

Technology, Wuhan 430074, P. R. China

[#]Department of Electronics, Peking University, Beijing 100871, P. R. China

[§]Department of Chemistry, University of Pennsylvania, Philadelphia, PA 19104, USA

[&]Department of Materials Science and Engineering, University of Pennsylvania, PA 19104,

USA.

*E-mail: shyliu@hust.edu.cn

*E-mail: dmj@seas.upenn.edu

**Raman Spectra of s-SWCNT**

Figure S1(a,b) shows the measurement position of Raman spectrometer and the typical Raman spectrum of s-SWCNT film is plotted in Figure S1(c,d). The diameter of CNT ($d_t$) can be evaluated from the frequencies of Radial Breathing Mode (RBM) by $\omega_{RBM}$–$d_t$ relation[1]

$$\omega_{RBM} = A/d_t + B \qquad (S1)$$

For the CNT on $SiO_2$/Si substrate, $A = 235$ cm$^{-1}$ and $B = 5.5$ cm$^{-1}$.[2] As illustrated in Figure S2(e), the RBM can be treated as a combination of two Lorentzian peaks, and the central frequencies for these two peaks are $\omega_{RBM1} \approx 151$ cm$^{-1}$ and $\omega_{RBM2} \approx 157$ cm$^{-1}$ (Table S1). This means that our CNT film samples mainly contain two different diameters ($d_t$) of s-SWCNT. Taking the fitting $\omega_{RBM}$ into Equation S1, the diameters can be calculated, the values are $d_{t1} \approx 1.62$ nm and $d_{t2} \approx 1.55$ nm .

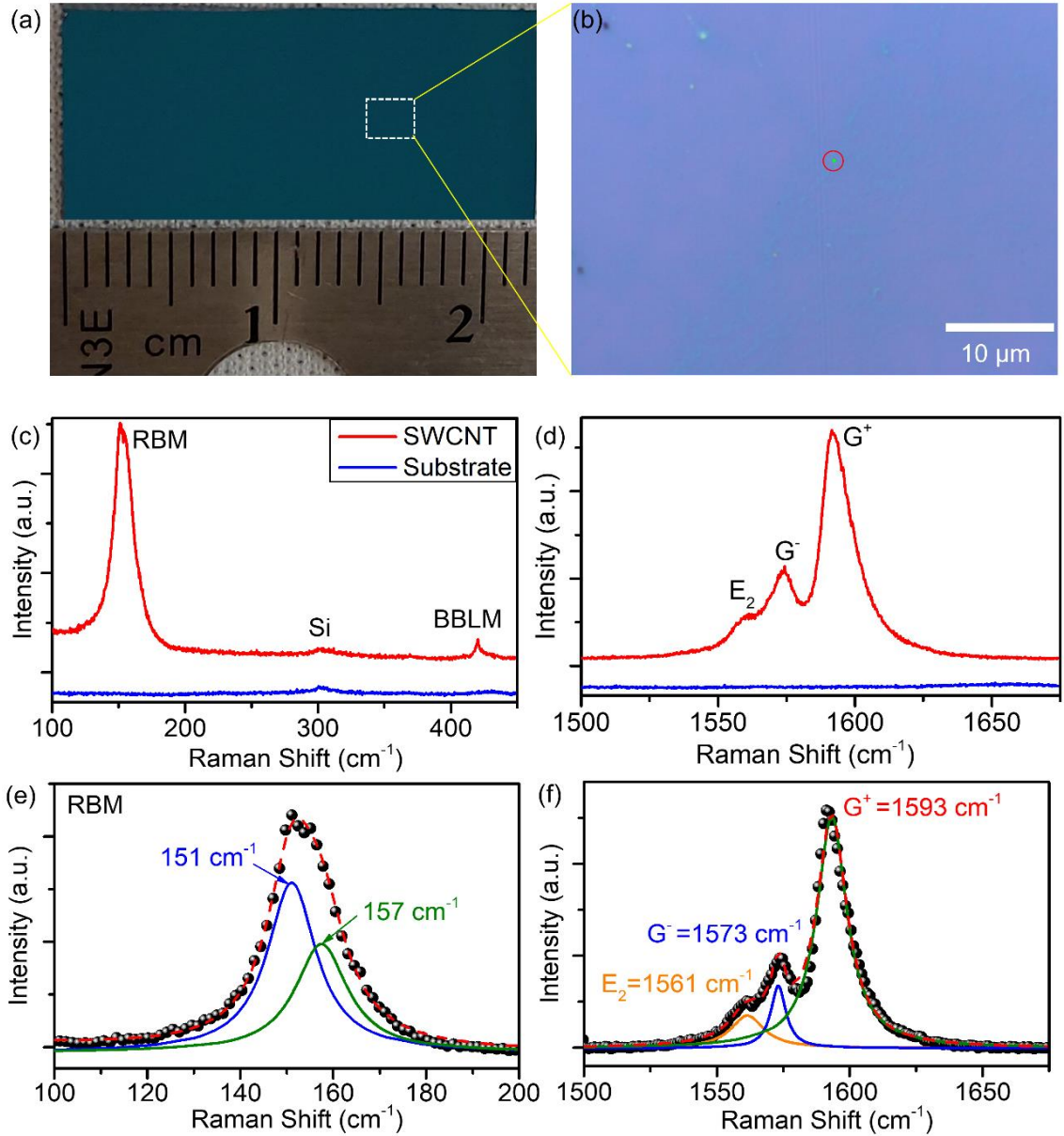

**Figure S1.** Raman spectra of s-SWCNT film. (a,b) Raman measurement spot (Green dot surrounded by a red circle) in a piece of s-SWCNT coated Si/SiO2 wafer. (c) Radial breathing mode (RBM) and bundle breathing-like mode (BBLM) at ~ 420 cm$^{-1}$. (d) G$^+$ feature at ~1593 cm$^{-1}$ is due to the in-plane vibrations along the tube axis, and a diameter-dependent G$^-$ feature at ~1573 cm$^{-1}$ for in-plane vibrations along the circumferential direction, and weak E$_2$ feature at ~1561 cm$^{-1}$ is associated with the symmetry phonons. (e,f) Fitting the RBM, G modes, and E feature with Lorentzian peaks.

For a specific SWCNT, the diameter can be uniquely calculated from its chiral indices (*n*, *m*), the relation can be written as[1]

$$d_t' = \sqrt{3}a_{cc}(m^2 + mn + n^2)^{1/2}/\pi \qquad (S2)$$

Where, $d_t'$ means the calculated diameter, $a_{cc}$ = 1.42 Å is the nearest-neighbor C–C distance of graphene. By virtue of Equation S2, we can inversely assign the chiral index $(n, m)$ by minimizing the deviation between $d_t'$ and $d_t$. Figure S2 illustrates the relative errors between $d_t'$ and $d_t$ of CNT with different $(n, m)$ (here, only $n \geq m$ case is considered, red box triangle regions in Figure S2). As shown in Figure S2(a), when $(n, m)$ is (17, 6), the relative error between $d_t'$ and $d_{t1}$ can achieve the minimum 0.17%, which means the first kind of SWCNT is $S_1$ [$(2n + m)$ mod 3 = 1] type semiconducting SWCNT, and its family and chiral angle are 40 and 14.56°. Similarly, the chiral index of SWCNT with $d_{t2}$ = 1.55 nm is (16, 6) (Figure S2(b)). The relative error between $d_t'$ and $d_{t2}$ is about 0.51%. The SWCNT is $S_2$ [$(2n + m)$ mod 3 = 2] type with a 15.29° chiral angle, and it attributes to family 38. The diameters and families obtained here are also in agreement with the literature precedent.[2]

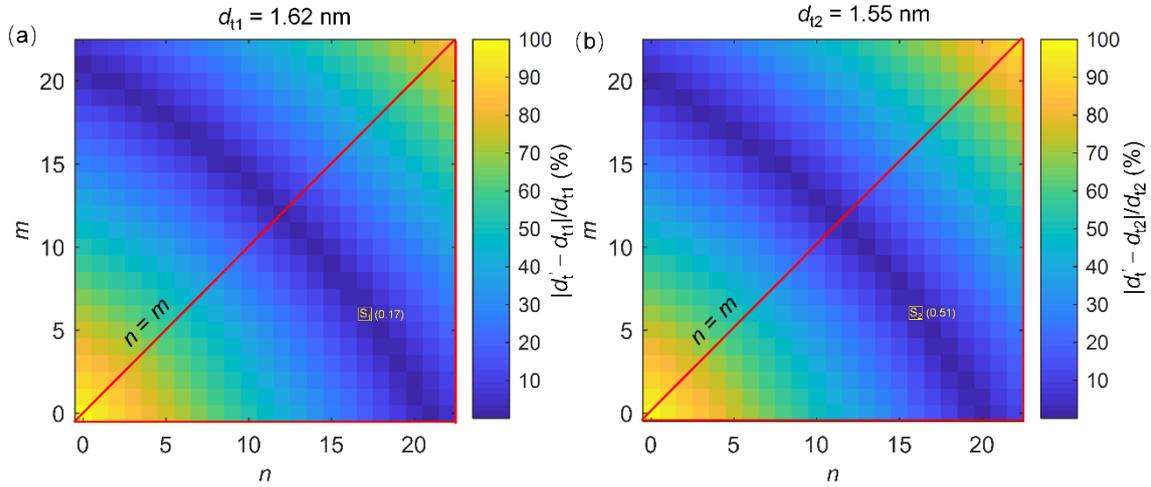

**Figure S2.** Relative errors between $d_t'$ and $d_t$ for the SWCNT with different chiral indices $(n, m)$. Symbols $S_1$, and $S_2$ denote two-type semiconducting SWCNT, and the values in round brackets refer to the relative errors.

**Spectroscopic Ellipsometry of s-SWCNT**

Ellipsometry is a common optical characterization technique that can simultaneously determine the optical and geometrical parameters of film materials in a nondestructive manner by recording the polarization state changes of the light before and after the reflection (transmission) by the samples.[3–5] The polarization state changes are usually described by a pair of ellipsometric angles $[\Psi(\omega),\ \Delta(\omega)]$ ($\omega$: the angular frequency of light). The relation between $[\Psi(\omega),\ \Delta(\omega)]$ and polarized reflection coefficients $r_p$ and $r_s$ (the subscript p and s mean p-polarization and s-polarization) can be expressed as[3]

$$\tan\Psi \cdot \exp(\mathrm{i}\Delta) = r_p / r_s,\qquad\qquad\text{(S3)}$$

where, $\tan\Psi$ and $\Delta$ represent amplitude ratio and phase difference between the p- and s-components of the polarized light, and $r_p$ and $r_s$ can be written as

$$r_p / r_s = (\mathbf{E}_{rp} / \mathbf{E}_{ip}) / (\mathbf{E}_{rs} / \mathbf{E}_{is}).\qquad\qquad\text{(S4)}$$

$\mathbf{E}_{ip}$ and $\mathbf{E}_{rs}$ represent the electric field vectors of incident p-polarization and reflected s-polarization (Figure S3).

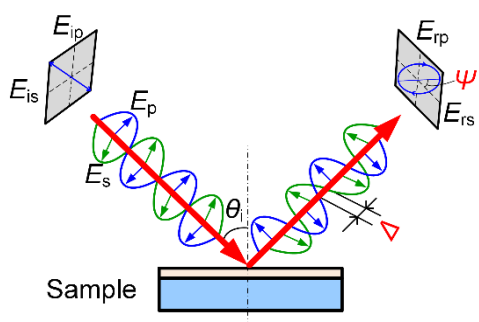

**Figure S3.** Basic principles of ellipsometry.

The spectroscopic ellipsometer used in our experiment is M-2000 type spectroscopic ellipsometer made by J.A. Woollam Company.[6] The ellipsometer covers a wavelength range of 371 nm–1687 nm (0.73 eV –3.34 eV). In the experiments, multi-incidence measurement mode is adopted to measure the ellipsometric spectra of SWCNT, which is conductive to reduce correlations among the fitting parameters in the subsequent analysis. The incident angles are 60°, 65°, and 70°. The measured multiangle ellipsometric spectra of s-SWCNT film under different gate voltage ($V_G$) are plotted in Figure S4.

Ellipsometry is a model-based technique. Two models (the optical model and the dielectric function model) need to be established to analyze the measured ellipsometric spectra. The optical model describes the light propagation through the optical stack of the sample calculated by the transfer matrix method (TMM). The dielectric function model gives the dispersive properties of the materials. For the s-SWCNT specimen, the optical model is a vertical stacking layered structure, including the ambient air, the s-SWCNT film, and the $SiO_2$/Si substrate (inset in Figure 3(a)). A combined oscillator model containing three Lorentz oscillators and two Gaussian oscillators is used to embody the dielectric responses of the s-SWCNT film. The dielectric function model is established by analyzing the s-SWCNT/sapphire specimen in which the ellipsometric spectrum features are relatively easy to fit. Based on the above two models, the theoretical ellipsometric spectra of the s-SWCNT film can be calculated, and then the measured ellipsometric spectra can be fitted by theoretical ones and extract the target optical and geometrical parameters. Figure S5 illustrates the measured and best-fitting ellipsometric spectra of the s-SWCNT film under different $V_G$. The excellent fitting goodness (the root-mean-square error (RMSE) is about 4.83) suggests that the ellipsometric models constructed by us are reasonable and the analysis process is reliable.

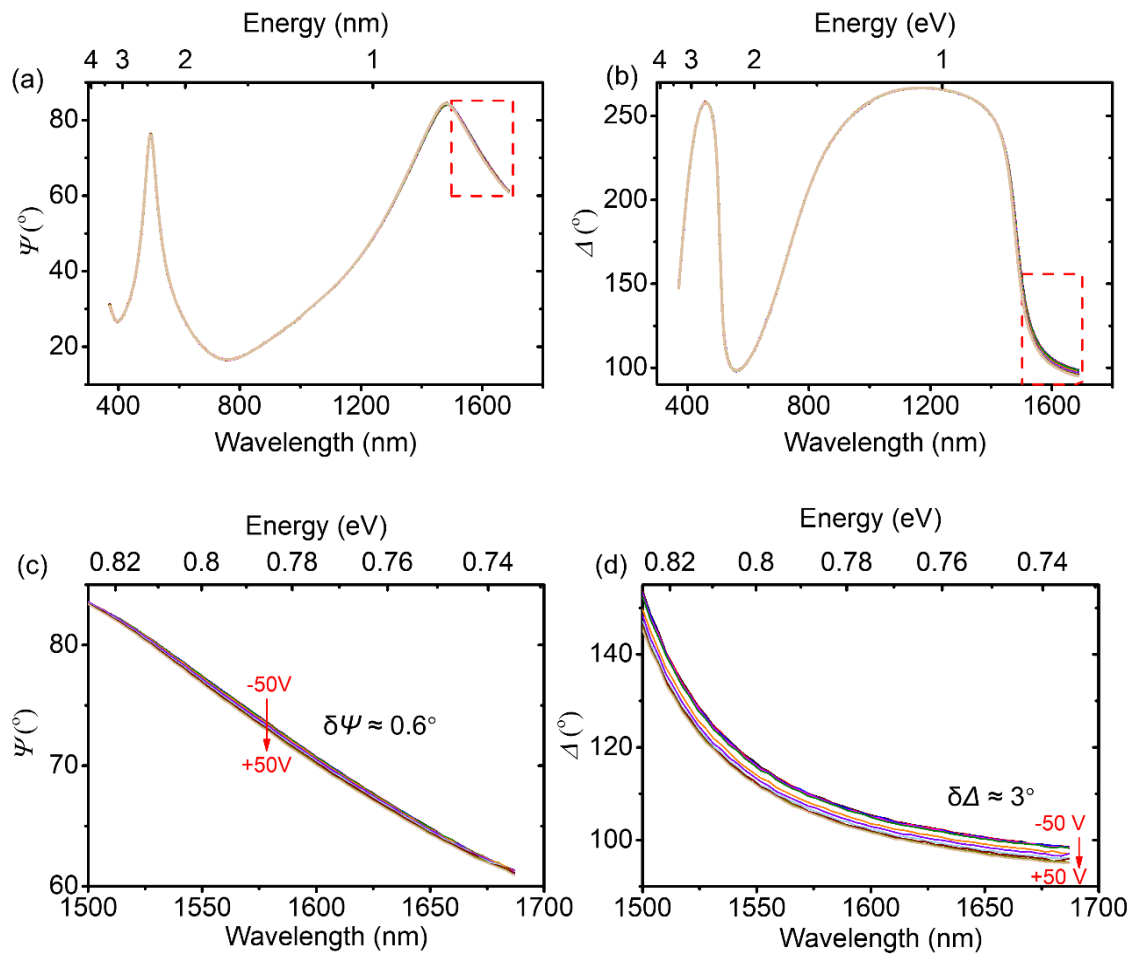

**Figure S4.** (a,b) Measured ellipsometric spectra ($\Psi(\omega)$, $\Delta(\omega)$) of SWCNT under different $V_G$. (c,d) Magnified views of the red dash rectangles in (a) and (b). The voltage-dependency of $\Delta(\omega)$ is stronger than that of $\Psi(\omega)$, suggesting that the $n$ of SWCNT is more sensitive to the $V_G$. The largest difference between $\Psi_{-50V}$ and $\Psi_{+50V}$ is about 0.6°, while the difference between $\Delta_{-50V}$ and $\Delta_{+50V}$ is as high as 3°.

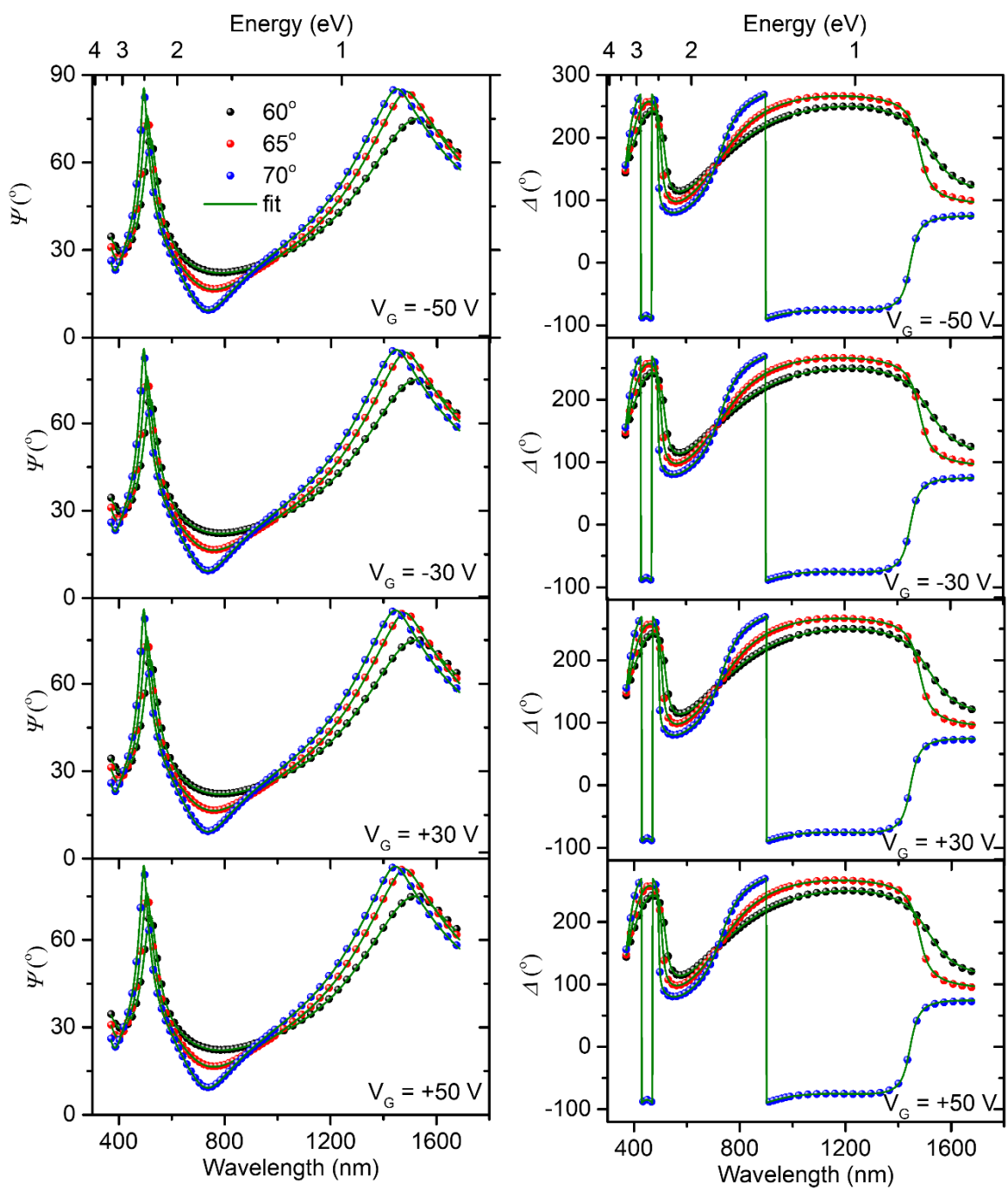

**Figure S5.** Measured and best fitting ellipsometric spectra of s-SWCNT film under different V$_G$.

## Complex refractive indices of Arc discharge and HiPco s-SWCNT films

The suspended s-SWCNTs in the chloroform solution used in this experiment are fabricated by Arc discharge method and their nominal diameters are about 1.4 nm–1.6 nm, and the average length is ~1.5 μm–2 μm. In order to investigate the optical constants of s-SWCNT with different diameters, we also prepared high-purity (>99%) HiPCO s-SWCNT films on the SiO$_2$/Si substrate by dip-coating method, and obtained its complex refractive index by SE measurement. The diameter of HiPCO s-SWCNT is 0.8 nm–1.1 nm, which is determined by similar Raman analysis method as provided in the above section. The average length of HiPCO s-SWCNTs is about 1.5 μm (roughly estimated from Figure S6(a)). As shown in Figure S6(a), the HiPco s-SWCNTs are obviously thinner than the Arc discharge s-SWCNTs (Figure 1(c)) and with lower root-mean-square surface roughness ($S_q$). One can also interpret that since the HiPCO s-SWCNTs are smaller in diameter, they attach themselves better to the substrate surface, forming a compact and flat film with fewer bundles.

Figure S6(b,c) shows the complex refractive indices of Arc discharge and HiPCO s-SWCNT film in an unbiased state (Table S3). Both the refractive index and the extinction coefficient of the HiPCO s-SWCNTs are larger than those of Arc discharge s-SWCNTs over the measured spectral range, except for the $\kappa$ spectra in the spectral region of ~1400 nm – 1690 nm. We infer this difference to the stronger exciton oscillator strengths in the SWCNTs with smaller diameters. Besides, the packing density of carbon nanotubes as well as the bundling may also affect the strength of optical constants since bundling results in screening/delocalization of excitons. Whereas, the precise contributions from these other factors are difficult to attain at this measurement due to material limitations. In the $\kappa$ spectrum of HiPCO s-SWCNTs, the S$_{11}$ and S$_{22}$ transition peaks are formed by a combination of multiple optical transitions, and the S$_{11}$ peaks can be entirely observed within our measured spectral

range, indicating that the optical bandgap of HiPCO s-SWCNT is larger than that of Arc discharge s-SWCNTs which have $S_{11}$ transition peak between 1600-1800 nm.[7] Since the optical density of states are all concentrated in the van hove singularities of the excitonic transitions,[8,9] it is expected that the nanotubes exhibit highest extinction values at or near $S_{11}$ transitions. This explains why the $\kappa$ value of Arc discharge s-SWCNT in the ~1400 nm – 1690 nm spectral range is higher as compared to HiPCO tubes. Both the $S_{11}$ and $S_{22}$ peaks of HiPCO s-SWCNTs appear at higher energies (shorter wavelengths) as compared with those of Arc discharge s-SWCNTs, consistent with the prior publications and explained as the stronger electron-hole interaction in smaller diameter tubes due to enhanced 1D quantum confinement.[10,11]

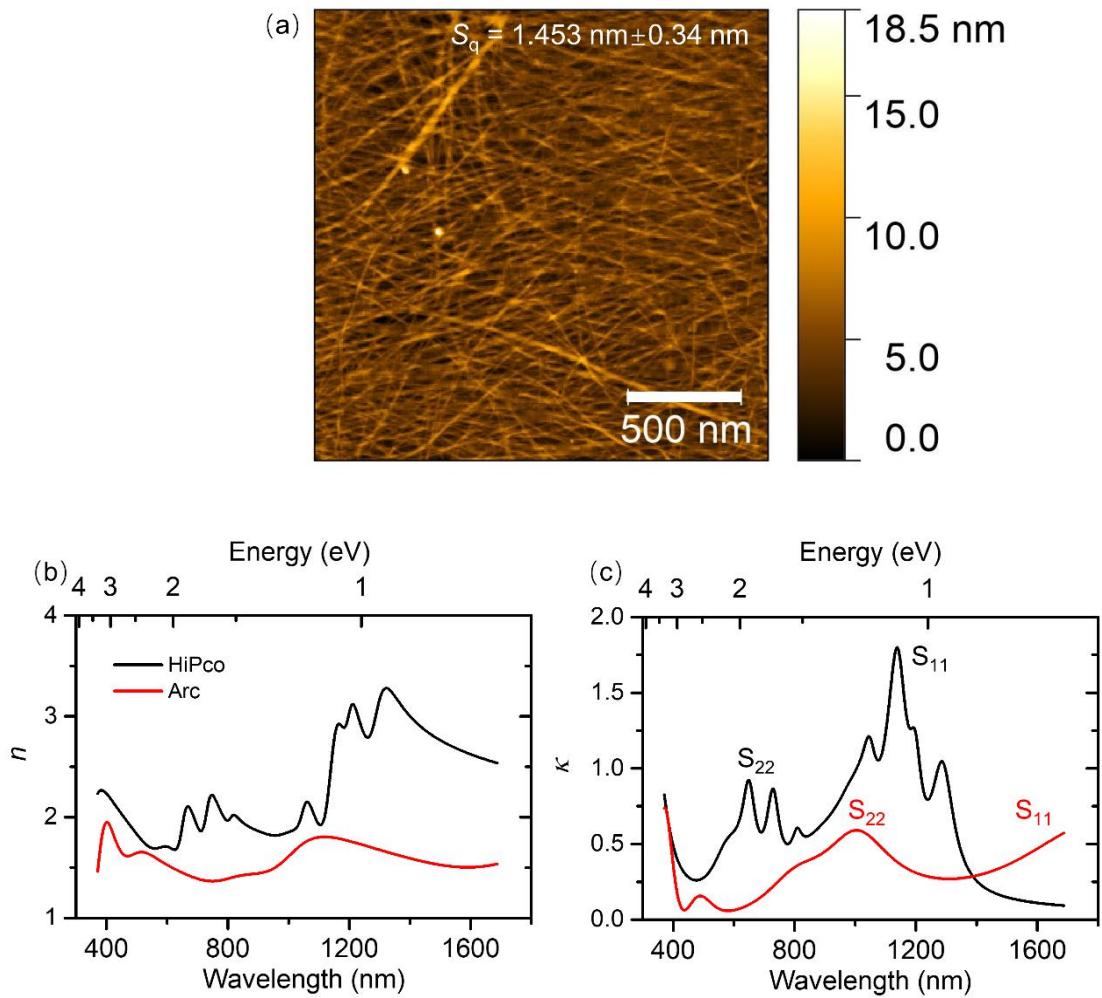

**Figure S6.** (a) The topographic image of HiPCO s-SWCNT film acquired using atomic force microscopy, the scan area is 2 μm × 2 μm. (b) Refractive indices *n* and (c) Extinction coefficients *κ* of Arc discharge and HiPCO s-SWCNT films.

## Near-infrared (IR) reflection phase modulator based on s-SWCNT film

Figure S7 shows optimized results of (1L-MoS$_2$/s-SWCNT)$_N$/MoS$_2$/Au superlattice reflection phase modulator, where *N* means the number of unit-cell. As show in Figure S7(a), when the thickness of s-SWCNT is fixed at 2.04 nm, a maximum average gate-tunable phase change −2.37° can be attained as the thickness of bottom MoS$_2$ and *N* are 72 nm and 5. It is worth noting that increasing the number of unit-cell (*N*) could improve the gate-tunable phase change. Here, considering the manufacturing complexity in reality, we only shown the results of (1L-MoS$_2$/s-SWCNT)$_5$/MoS$_2$/Au. Additionally, we found that increasing the layer number of MoS$_2$ in the unit-cell will cause a slight attenuation of the changing range of the average reflection phase (Figure S7(d)). Here, we want to make a simple interpretation about the setting of optical constants in the simulations. Actually, we assume that the isotropic complex refractive index shown in Figure 2(c,d) is equal to the in-plane refractive index. The isotropic complex refractive index of the ultrathin s-SWCNT film (2.04 nm) can largely reflect the in-plane optical responses due to the extremely weak out-of-plane contribution. Furthermore, the s-SWCNT we prepared exhibits random distribution on the substrate surface (Figure 1(b,c)), thus, it can be equivalently regarded as a uniaxial material with the dielectric tensor diag $\left[\varepsilon_o, \varepsilon_o, \varepsilon_e\right]$ ($\varepsilon_o$=$N_o^2$ and $\varepsilon_e$=$N_e^2$ refer to in-plane and out-of-plane dielectric functions),[12] and the optical axis is perpendicular to the s-SWCNT film surface.[13] Therefore, $N_o$ dominates the optical responses of s-SWCNT when the light is nominal incidence. To sum up, the isotropic complex refractive index of 2.04 nm s-SWCNT could be regarded as the in-plane refractive index and can be adopted in all simulations of this work.

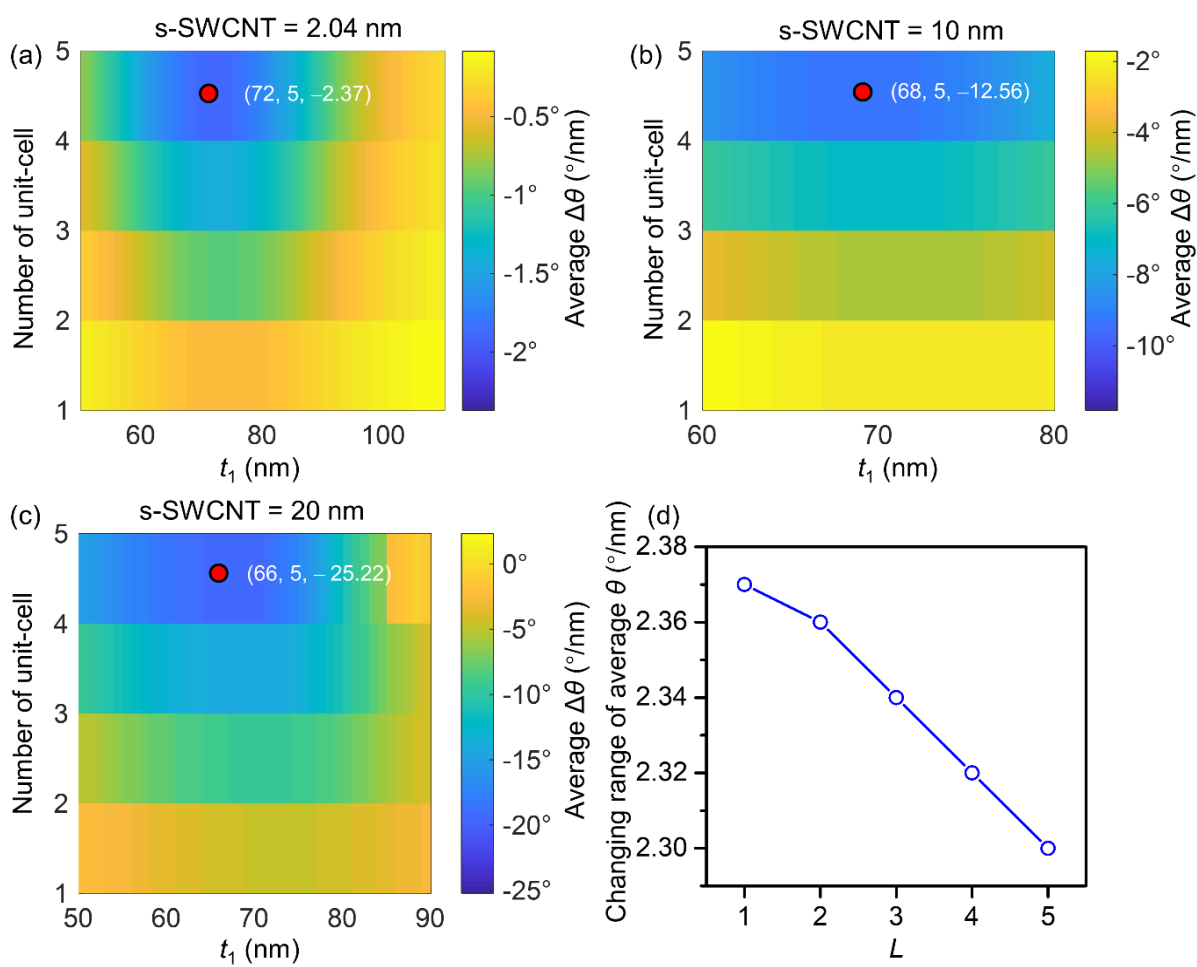

**Figure S7.** (a–c) Optimized results of (1L-MoS₂/s-SWCNT)₅/MoS₂/Au reflection phase modulator when the thicknesses of s-SWCNT are 2.04 nm, 10 nm, and 20 nm. The three white numbers in the round brackets indicate the best bottom MoS₂ thickness, the best number of the unit-cell and the maximum average reflection phase changes. (d) The maximum changing range of the average reflection phase vs. the layer number ($L$) of MoS₂ in the unit-cell.

**Near-IR absorber based on s-SWCNT**

A near-IR absorber with the (1L-MoS₂/s-SWCNT)$_N$/MoS₂/Au superlattice structure is designed, and Figure S8(a–c) illustrate the optimized results. The optical constants of s-SWCNT we used in this simulation are without applied voltage (0V). The maximum average absorptance increases with the thickness of s-SWCNT, and at the same time the thickness of bottom MoS$_2$ will decrease. As shown in Figure S8(d), when the thickness of s-SWCNT is set as 20 nm, the optimized absorptance of the absorber will be larger than 0.65 in the entire concerned spectral region 1300–1687 nm. The largest absorptance can be close to 0.9.

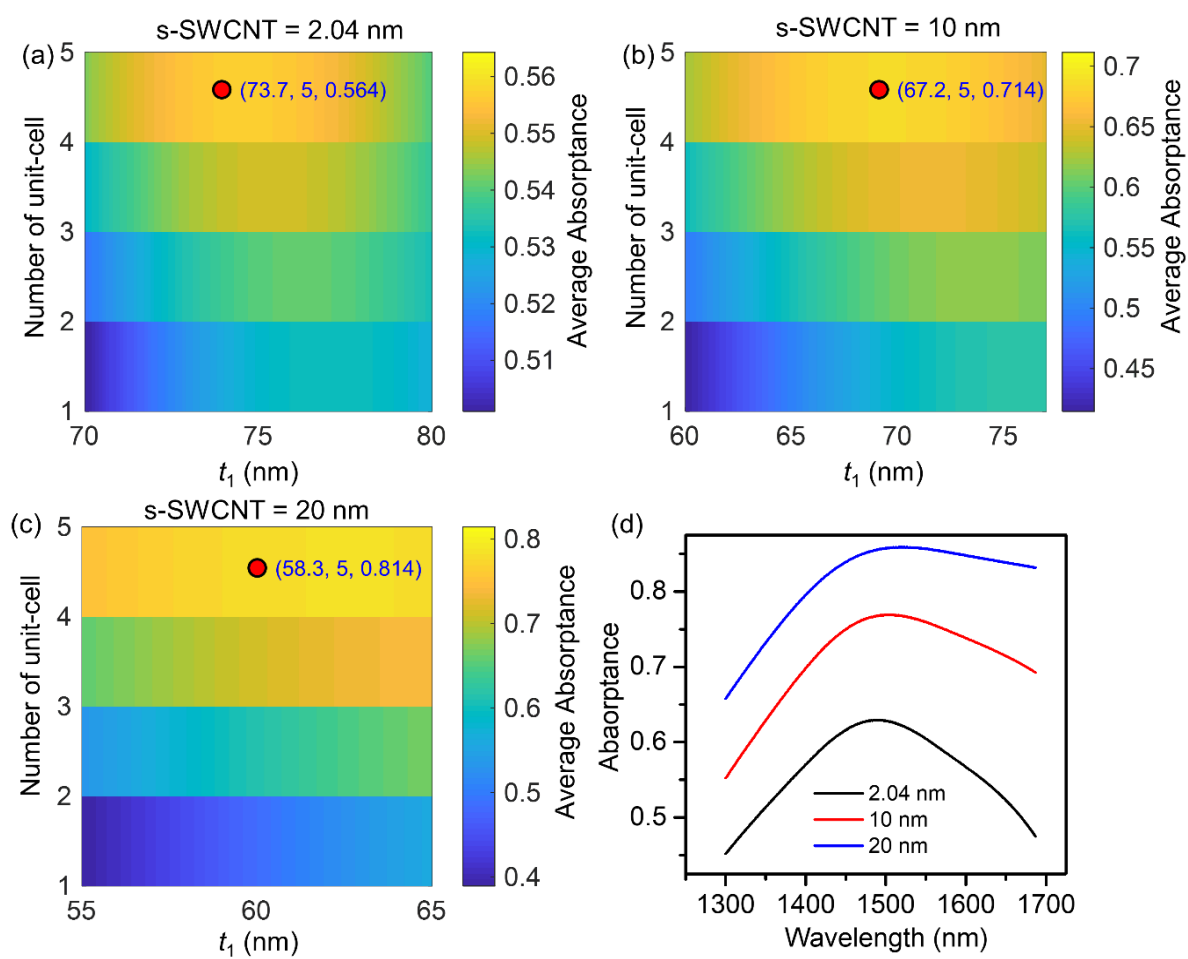

**Figure S8.** (a–c) Optimized results of IR absorber based on the (1L-MoS$_2$/s-SWCNT)$_5$/MoS$_2$/Au superlattice structure. The three white numbers in the round brackets indicate the best bottom MoS$_2$ thickness, the best number of the unit-cell and the maximum average absorptance. (d) The optimized absorptances of (1L-MoS$_2$/s-SWCNT)$_5$/MoS$_2$/Au when the thicknesses of s-SWCNT in the unit-cell are set as 2.04 nm, 10 nm, and 20 nm, respectively.

**Table S1.** Typical Raman modes of s-SWCNT

| Raman mode | $\omega$ (cm$^{-1}$) |
|---|---|
| RBM | $\omega_{\text{RBM1}} = 151$<br>$\omega_{\text{RBM2}} = 157$ |
| BBLM | 420 |
| D | 1327 |
| G$^-$ | 1574 |
| G$^+$ | 1591 |

**Table S2.** $\Delta n$ and $\Delta \kappa$ of s-SWCNT, traditional semiconductors, and Quantum Well (QW) structures

| s-SWCNT | | | | | | |
|---|---|---|---|---|---|---|
| Electric field (MV/m) | $\Delta n$ (1300 nm) | $\Delta n$ (1550 nm) | $\Delta n$ (1600 nm) | $\Delta\alpha$ (1300 nm) (cm$^{-1}$) | $\Delta\alpha$ (1550 nm) (cm$^{-1}$) | $\Delta\alpha$ (1600 nm) (cm$^{-1}$) |
| −154.08 | 0.042 | 0.068 | 0.072 | −634.20 | −1428.47 | −1674.23 |
| −123.27 | 0.040 | 0.065 | 0.069 | −610.17 | −1372.01 | −1607.55 |
| −92.36 | 0.037 | 0.060 | 0.063 | −569.42 | −1274.96 | −1492.88 |
| −61.63 | 0.033 | 0.053 | 0.056 | −500.67 | −1124.86 | −1317.51 |
| −30.82 | 0.029 | 0.047 | 0.049 | −448.05 | −1003.29 | −1174.48 |
| 0 | 0 | 0 | 0 | 0 | 0 | 0 |
| 30.82 | −0.020 | −0.031 | −0.033 | 309.86 | 700.56 | 819.48 |
| 61.63 | −0.040 | −0.064 | −0.067 | 636.19 | 1471.28 | 1724.89 |
| 92.36 | −0.054 | −0.089 | −0.095 | 869.94 | 2101.28 | 2477.15 |
| 123.27 | −0.065 | −0.11 | −0.12 | 1045.17 | 2670.71 | 3171.67 |
| 154.08 | −0.069 | −0.13 | −0.13 | 1125.48 | 3032.03 | 3625.87 |

| Si [14,15] | | | InP [16] | | |
|---|---|---|---|---|---|
| Electric field (MV/m) | $\Delta n$ (1070 nm) | $\Delta\alpha$ (1070 nm) (cm$^{-1}$) | Electric field (MV/m) | $\Delta n$ (943 nm) | $\Delta\alpha$ (943 nm) (cm$^{-1}$) |
| 4 | | 0.0525 | 0.5 | 2.94E-05 | 1.51 |
| 7.5 | 5.12E-06 | 0.50 | | | |
| 8.5 | 8.56E-06 | 1.07 | 1 | 9.35E-05 | 7.00 |
| 10 | 1.32E-05 | 1.61 | 1.5 | 2.13E-04 | 18.25 |

| Ge [17] | | | GaAs [17] | | |
|---|---|---|---|---|---|
| Electric field (MV/m) | $\Delta n$ (1851 nm) | $\Delta\alpha$ (1851 nm) (cm$^{-1}$) | Electric field (MV/m) | $\Delta n$ (860 nm) | $\Delta\alpha$ (860 nm) (cm$^{-1}$) |
| 5 | −0.013 | 748.91 | 5 | −0.010 | 984.13 |
| 10 | −0.019 | 1006.38 | 10 | −0.014 | 1274.38 |
| 15 | −0.022 | 1162.71 | 15 | −0.016 | 1473.92 |
| 20 | −0.025 | 1263.86 | 20 | −0.018 | 1609.98 |

| Monolayer Graphene [18] | | | | | Ge/SiGe QW [19] | |
|---|---|---|---|---|---|---|
| Electric field (MV/m) | $\Delta n$ (1300 nm) | $\Delta n$ (1550 nm) | $\Delta\alpha$ (1300 nm) (cm$^{-1}$) | $\Delta\alpha$ (1550 nm) (cm$^{-1}$) | Electric field (MV/m) | $\Delta n$ (1430 nm) |
| 0 | 0 | 0 | 0 | 0 | 1 | 6.00E-06 |
| 33.33 | 0.070 | 0.039 | −15323.08 | −10178.63 | 2 | 5.89E-04 |
| 56.67 | 0.0097 | –0.036 | –7052.53 | –2966.10 | 3 | 1.00E-03 |
| 93.33 | 0.0043 | –0.019 | –9308.40 | –5616.69 | 4 | 1.25E-03 |
| 133.33 | 0.085 | –0.026 | –6861.75 | –1180.75 | 5 | 4.44E-04 |
| 180 | 0.0055 | –0.078 | –2943.35 | 1570.81 | 10 | 1.74E-03 |
| 236.67 | 0.070 | –0.091 | –2974.66 | 9101.02 | | |

**InGaAsN QW[20]**

| Electric field (MV/m) | $\Delta n$ (1179 nm) | Electric field (MV/m) | $\Delta \alpha$ (1179 nm) |
|---|---|---|---|
| 3.71 | –0.0022 | 3.71 | 482.76 |
| 11.14 | –0.022 | 11.14 | 1753.45 |
| 26 | –0.046 | 14.86 | 1530.22 |
| | | 18.57 | 774.67 |
| | | 22.29 | 1.95 |
| | | 26 | –358.65 |

**Table S3.** $n$ and $\kappa$ of Arc discharge and HiPco s-SWCNT film on SiO$_2$/Si substrate

| Wavelength (nm) | $n$-Arc | $\kappa$-Arc | $n$-HiPco | $\kappa$-HiPco |
|---|---|---|---|---|
| 370.96 | 1.46 | 0.73 | 2.24 | 0.83 |
| 372.54 | 1.51 | 0.74 | 2.24 | 0.80 |
| 374.12 | 1.55 | 0.74 | 2.25 | 0.78 |
| 375.71 | 1.60 | 0.73 | 2.26 | 0.76 |
| 377.29 | 1.64 | 0.72 | 2.26 | 0.73 |
| 378.87 | 1.68 | 0.71 | 2.26 | 0.71 |
| 380.45 | 1.72 | 0.69 | 2.27 | 0.69 |
| 382.04 | 1.75 | 0.68 | 2.27 | 0.67 |
| 383.62 | 1.79 | 0.65 | 2.27 | 0.65 |
| 385.20 | 1.81 | 0.63 | 2.27 | 0.63 |
| 386.79 | 1.84 | 0.60 | 2.27 | 0.62 |
| 388.37 | 1.87 | 0.58 | 2.26 | 0.60 |
| 389.95 | 1.89 | 0.55 | 2.26 | 0.58 |
| 391.54 | 1.90 | 0.52 | 2.26 | 0.57 |
| 393.12 | 1.92 | 0.49 | 2.25 | 0.55 |
| 394.70 | 1.93 | 0.46 | 2.25 | 0.53 |
| 396.29 | 1.94 | 0.43 | 2.25 | 0.52 |
| 397.87 | 1.94 | 0.40 | 2.24 | 0.51 |
| 399.46 | 1.95 | 0.38 | 2.24 | 0.49 |
| 401.04 | 1.95 | 0.35 | 2.23 | 0.48 |
| 402.62 | 1.95 | 0.32 | 2.22 | 0.47 |
| 404.21 | 1.95 | 0.30 | 2.22 | 0.46 |
| 405.79 | 1.94 | 0.27 | 2.21 | 0.45 |
| 407.38 | 1.94 | 0.25 | 2.21 | 0.44 |
| 408.96 | 1.93 | 0.23 | 2.20 | 0.43 |
| 410.55 | 1.92 | 0.21 | 2.19 | 0.42 |
| 412.13 | 1.91 | 0.19 | 2.19 | 0.41 |
| 413.72 | 1.90 | 0.17 | 2.18 | 0.40 |
| 415.30 | 1.89 | 0.15 | 2.18 | 0.39 |
| 416.89 | 1.88 | 0.14 | 2.17 | 0.38 |
| 418.48 | 1.86 | 0.13 | 2.16 | 0.38 |
| 420.06 | 1.85 | 0.11 | 2.16 | 0.37 |
| 421.65 | 1.84 | 0.10 | 2.15 | 0.36 |
| 423.23 | 1.82 | 0.10 | 2.14 | 0.35 |
| 424.82 | 1.81 | 0.09 | 2.14 | 0.35 |
| 426.41 | 1.80 | 0.08 | 2.13 | 0.34 |
| 427.99 | 1.78 | 0.08 | 2.12 | 0.34 |
| 429.58 | 1.77 | 0.07 | 2.12 | 0.33 |
| 431.17 | 1.76 | 0.07 | 2.11 | 0.33 |
| 432.75 | 1.75 | 0.07 | 2.10 | 0.32 |
| 434.34 | 1.74 | 0.06 | 2.10 | 0.32 |
| 435.93 | 1.73 | 0.06 | 2.09 | 0.31 |
| 437.51 | 1.71 | 0.06 | 2.08 | 0.31 |
| 439.10 | 1.70 | 0.06 | 2.08 | 0.30 |
| 440.69 | 1.69 | 0.07 | 2.07 | 0.30 |
| 442.28 | 1.69 | 0.07 | 2.06 | 0.30 |
| 443.86 | 1.68 | 0.07 | 2.06 | 0.29 |
| 445.45 | 1.67 | 0.07 | 2.05 | 0.29 |
| 447.04 | 1.66 | 0.07 | 2.04 | 0.29 |
| 448.63 | 1.65 | 0.08 | 2.04 | 0.28 |
| 450.22 | 1.65 | 0.08 | 2.03 | 0.28 |
| 451.80 | 1.64 | 0.09 | 2.02 | 0.28 |
| 453.39 | 1.64 | 0.09 | 2.02 | 0.28 |
| 454.98 | 1.63 | 0.09 | 2.01 | 0.27 |
| 456.57 | 1.63 | 0.10 | 2.00 | 0.27 |
| 458.16 | 1.62 | 0.10 | 2.00 | 0.27 |
| 459.75 | 1.62 | 0.11 | 1.99 | 0.27 |
| 461.33 | 1.62 | 0.11 | 1.98 | 0.27 |
| 462.92 | 1.62 | 0.12 | 1.98 | 0.27 |
| 464.51 | 1.62 | 0.12 | 1.97 | 0.26 |
| 466.10 | 1.61 | 0.13 | 1.96 | 0.26 |
| 467.69 | 1.61 | 0.13 | 1.96 | 0.26 |
| 469.28 | 1.61 | 0.13 | 1.95 | 0.26 |
| 470.87 | 1.61 | 0.14 | 1.95 | 0.26 |
| 472.46 | 1.61 | 0.14 | 1.94 | 0.26 |
| 474.05 | 1.62 | 0.14 | 1.93 | 0.26 |
| 475.64 | 1.62 | 0.15 | 1.93 | 0.26 |
| 477.23 | 1.62 | 0.15 | 1.92 | 0.26 |
| 478.82 | 1.62 | 0.15 | 1.91 | 0.26 |
| 480.41 | 1.62 | 0.15 | 1.91 | 0.26 |
| 482.00 | 1.62 | 0.15 | 1.90 | 0.26 |
| 483.59 | 1.63 | 0.16 | 1.90 | 0.26 |
| 485.18 | 1.63 | 0.16 | 1.89 | 0.26 |
| 486.77 | 1.63 | 0.16 | 1.88 | 0.26 |
| 488.36 | 1.63 | 0.16 | 1.88 | 0.26 |
| 489.95 | 1.63 | 0.16 | 1.87 | 0.27 |
| 491.54 | 1.64 | 0.16 | 1.86 | 0.27 |
| 493.13 | 1.64 | 0.16 | 1.86 | 0.27 |
| 494.72 | 1.64 | 0.16 | 1.85 | 0.27 |
| 496.31 | 1.64 | 0.15 | 1.85 | 0.27 |
| 497.90 | 1.64 | 0.15 | 1.84 | 0.27 |
| 499.49 | 1.65 | 0.15 | 1.83 | 0.28 |
| 501.08 | 1.65 | 0.15 | 1.83 | 0.28 |
| 502.67 | 1.65 | 0.15 | 1.82 | 0.28 |
| 504.26 | 1.65 | 0.15 | 1.82 | 0.28 |
| 505.85 | 1.65 | 0.14 | 1.81 | 0.29 |
| 507.44 | 1.65 | 0.14 | 1.80 | 0.29 |
| 509.04 | 1.65 | 0.14 | 1.80 | 0.29 |
| 510.63 | 1.65 | 0.14 | 1.79 | 0.30 |
| 512.22 | 1.65 | 0.13 | 1.79 | 0.30 |
| 513.81 | 1.65 | 0.13 | 1.78 | 0.30 |
| 515.40 | 1.65 | 0.13 | 1.78 | 0.31 |
| 516.99 | 1.65 | 0.12 | 1.77 | 0.31 |
| 518.58 | 1.65 | 0.12 | 1.77 | 0.31 |
| 520.18 | 1.65 | 0.12 | 1.76 | 0.32 |
| 521.77 | 1.65 | 0.12 | 1.76 | 0.32 |
| 523.36 | 1.65 | 0.11 | 1.75 | 0.33 |
| 524.95 | 1.65 | 0.11 | 1.75 | 0.33 |
| 526.54 | 1.65 | 0.11 | 1.74 | 0.34 |
| 528.13 | 1.65 | 0.10 | 1.74 | 0.34 |
| 529.73 | 1.65 | 0.10 | 1.73 | 0.35 |
| 531.32 | 1.65 | 0.10 | 1.73 | 0.35 |
| 532.91 | 1.64 | 0.10 | 1.72 | 0.36 |
| 534.50 | 1.64 | 0.09 | 1.72 | 0.37 |
| 536.09 | 1.64 | 0.09 | 1.72 | 0.37 |
| 537.69 | 1.64 | 0.09 | 1.71 | 0.38 |
| 539.28 | 1.64 | 0.09 | 1.71 | 0.38 |
| 540.87 | 1.63 | 0.08 | 1.71 | 0.39 |
| 542.46 | 1.63 | 0.08 | 1.70 | 0.40 |
| 544.06 | 1.63 | 0.08 | 1.70 | 0.40 |
| 545.65 | 1.63 | 0.08 | 1.70 | 0.41 |
| 547.24 | 1.62 | 0.08 | 1.70 | 0.42 |
| 548.83 | 1.62 | 0.07 | 1.69 | 0.42 |
| 550.43 | 1.62 | 0.07 | 1.69 | 0.43 |
| 552.02 | 1.61 | 0.07 | 1.69 | 0.44 |
| 553.61 | 1.61 | 0.07 | 1.69 | 0.44 |
| 555.20 | 1.61 | 0.07 | 1.69 | 0.45 |
| 556.80 | 1.61 | 0.07 | 1.69 | 0.46 |
| 558.39 | 1.60 | 0.07 | 1.69 | 0.46 |
| 559.98 | 1.60 | 0.06 | 1.69 | 0.47 |
| 561.58 | 1.60 | 0.06 | 1.69 | 0.48 |
| 563.17 | 1.59 | 0.06 | 1.69 | 0.48 |
| 564.76 | 1.59 | 0.06 | 1.69 | 0.49 |
| 566.36 | 1.59 | 0.06 | 1.69 | 0.50 |
| 567.95 | 1.59 | 0.06 | 1.69 | 0.50 |
| 569.54 | 1.58 | 0.06 | 1.69 | 0.51 |
| 571.13 | 1.58 | 0.06 | 1.69 | 0.51 |
| 572.73 | 1.58 | 0.06 | 1.70 | 0.52 |
| 574.32 | 1.57 | 0.06 | 1.70 | 0.52 |
| 575.91 | 1.57 | 0.06 | 1.70 | 0.53 |
| 577.51 | 1.57 | 0.06 | 1.70 | 0.53 |
| 579.10 | 1.56 | 0.06 | 1.70 | 0.54 |
| 580.69 | 1.56 | 0.06 | 1.70 | 0.54 |
| 582.29 | 1.56 | 0.06 | 1.70 | 0.55 |
| 583.88 | 1.55 | 0.06 | 1.71 | 0.55 |
| 585.47 | 1.55 | 0.06 | 1.71 | 0.55 |
| 587.07 | 1.55 | 0.06 | 1.71 | 0.56 |
| 588.66 | 1.55 | 0.06 | 1.71 | 0.56 |
| 590.25 | 1.54 | 0.06 | 1.71 | 0.56 |
| 591.85 | 1.54 | 0.06 | 1.71 | 0.57 |
| 593.44 | 1.54 | 0.06 | 1.71 | 0.57 |
| 595.03 | 1.53 | 0.06 | 1.71 | 0.57 |
| 596.63 | 1.53 | 0.06 | 1.71 | 0.58 |
| 598.22 | 1.53 | 0.06 | 1.71 | 0.58 |
| 599.81 | 1.53 | 0.06 | 1.71 | 0.58 |
| 601.41 | 1.52 | 0.06 | 1.71 | 0.59 |
| 603.00 | 1.52 | 0.06 | 1.70 | 0.59 |
| 604.60 | 1.52 | 0.06 | 1.70 | 0.60 |
| 606.19 | 1.52 | 0.07 | 1.70 | 0.60 |
| 607.78 | 1.51 | 0.07 | 1.70 | 0.61 |
| 609.38 | 1.51 | 0.07 | 1.70 | 0.61 |
| 610.97 | 1.51 | 0.07 | 1.69 | 0.62 |
| 612.56 | 1.51 | 0.07 | 1.69 | 0.63 |
| 614.16 | 1.50 | 0.07 | 1.69 | 0.63 |
| 615.75 | 1.50 | 0.07 | 1.69 | 0.64 |
| 617.34 | 1.50 | 0.07 | 1.68 | 0.65 |
| 618.94 | 1.50 | 0.07 | 1.68 | 0.66 |
| 620.53 | 1.49 | 0.07 | 1.68 | 0.67 |
| 622.13 | 1.49 | 0.07 | 1.68 | 0.69 |
| 623.72 | 1.49 | 0.08 | 1.68 | 0.70 |
| 625.31 | 1.49 | 0.08 | 1.68 | 0.71 |
| 626.91 | 1.48 | 0.08 | 1.68 | 0.73 |
| 628.50 | 1.48 | 0.08 | 1.68 | 0.74 |
| 630.09 | 1.48 | 0.08 | 1.68 | 0.76 |
| 631.69 | 1.48 | 0.08 | 1.69 | 0.78 |
| 633.28 | 1.47 | 0.08 | 1.70 | 0.80 |
| 634.87 | 1.47 | 0.08 | 1.71 | 0.82 |
| 636.47 | 1.47 | 0.09 | 1.72 | 0.83 |
| 638.06 | 1.47 | 0.09 | 1.74 | 0.85 |
| 639.65 | 1.46 | 0.09 | 1.75 | 0.87 |
| 641.25 | 1.46 | 0.09 | 1.77 | 0.88 |
| 642.84 | 1.46 | 0.09 | 1.80 | 0.90 |
| 644.44 | 1.46 | 0.09 | 1.82 | 0.91 |
| 646.03 | 1.45 | 0.10 | 1.85 | 0.92 |
| 647.62 | 1.45 | 0.10 | 1.88 | 0.92 |
| 649.22 | 1.45 | 0.10 | 1.90 | 0.92 |
| 650.81 | 1.45 | 0.10 | 1.93 | 0.92 |
| 652.40 | 1.45 | 0.10 | 1.96 | 0.91 |
| 654.00 | 1.44 | 0.10 | 1.99 | 0.90 |
| 655.59 | 1.44 | 0.11 | 2.01 | 0.89 |
| 657.18 | 1.44 | 0.11 | 2.04 | 0.87 |
| 658.78 | 1.44 | 0.11 | 2.06 | 0.85 |
| 660.37 | 1.44 | 0.11 | 2.07 | 0.83 |
| 661.96 | 1.43 | 0.11 | 2.09 | 0.81 |
| 663.56 | 1.43 | 0.11 | 2.10 | 0.79 |
| 665.15 | 1.43 | 0.12 | 2.10 | 0.77 |
| 666.74 | 1.43 | 0.12 | 2.11 | 0.75 |
| 668.34 | 1.43 | 0.12 | 2.11 | 0.72 |
| 669.93 | 1.42 | 0.12 | 2.11 | 0.70 |
| 671.52 | 1.42 | 0.12 | 2.10 | 0.68 |
| 673.12 | 1.42 | 0.13 | 2.10 | 0.67 |
| 674.71 | 1.42 | 0.13 | 2.09 | 0.65 |
| 676.30 | 1.42 | 0.13 | 2.08 | 0.64 |
| 677.90 | 1.41 | 0.13 | 2.07 | 0.62 |
| 679.49 | 1.41 | 0.13 | 2.06 | 0.61 |
| 681.08 | 1.41 | 0.14 | 2.05 | 0.60 |
| 682.67 | 1.41 | 0.14 | 2.04 | 0.59 |
| 684.27 | 1.41 | 0.14 | 2.03 | 0.59 |
| 685.86 | 1.40 | 0.14 | 2.02 | 0.58 |
| 687.45 | 1.40 | 0.15 | 2.00 | 0.58 |
| 689.05 | 1.40 | 0.15 | 1.99 | 0.58 |
| 690.64 | 1.40 | 0.15 | 1.98 | 0.58 |
| 692.23 | 1.40 | 0.15 | 1.96 | 0.58 |
| 693.82 | 1.40 | 0.16 | 1.95 | 0.58 |
| 695.42 | 1.39 | 0.16 | 1.94 | 0.58 |
| 697.01 | 1.39 | 0.16 | 1.93 | 0.59 |
| 698.60 | 1.39 | 0.16 | 1.92 | 0.60 |
| 700.20 | 1.39 | 0.17 | 1.91 | 0.60 |
| 701.79 | 1.39 | 0.17 | 1.90 | 0.61 |
| 703.38 | 1.39 | 0.17 | 1.89 | 0.62 |
| 704.97 | 1.39 | 0.17 | 1.88 | 0.64 |
| 706.57 | 1.38 | 0.18 | 1.87 | 0.65 |
| 708.16 | 1.38 | 0.18 | 1.87 | 0.67 |
| 709.75 | 1.38 | 0.18 | 1.86 | 0.68 |
| 711.34 | 1.38 | 0.19 | 1.86 | 0.70 |
| 712.93 | 1.38 | 0.19 | 1.86 | 0.72 |
| 714.53 | 1.38 | 0.19 | 1.87 | 0.74 |
| 716.12 | 1.38 | 0.19 | 1.87 | 0.76 |
| 717.71 | 1.38 | 0.20 | 1.88 | 0.78 |
| 719.30 | 1.37 | 0.20 | 1.90 | 0.80 |
| 720.89 | 1.37 | 0.20 | 1.91 | 0.82 |
| 722.49 | 1.37 | 0.21 | 1.93 | 0.83 |
| 724.08 | 1.37 | 0.21 | 1.95 | 0.85 |
| 725.67 | 1.37 | 0.21 | 1.98 | 0.86 |
| 727.26 | 1.37 | 0.22 | 2.01 | 0.86 |
| 728.85 | 1.37 | 0.22 | 2.03 | 0.87 |
| 730.44 | 1.37 | 0.22 | 2.06 | 0.86 |
| 732.04 | 1.37 | 0.22 | 2.09 | 0.86 |
| 733.63 | 1.37 | 0.23 | 2.12 | 0.85 |
| 735.22 | 1.37 | 0.23 | 2.14 | 0.83 |
| 736.81 | 1.37 | 0.23 | 2.16 | 0.81 |
| 738.40 | 1.37 | 0.24 | 2.18 | 0.80 |
| 739.99 | 1.37 | 0.24 | 2.19 | 0.77 |
| 741.58 | 1.37 | 0.24 | 2.20 | 0.75 |
| 743.17 | 1.37 | 0.25 | 2.21 | 0.73 |
| 744.77 | 1.37 | 0.25 | 2.22 | 0.71 |
| 746.36 | 1.37 | 0.25 | 2.22 | 0.69 |
| 747.95 | 1.37 | 0.26 | 2.22 | 0.66 |
| 749.54 | 1.37 | 0.26 | 2.22 | 0.64 |
| 751.13 | 1.37 | 0.26 | 2.21 | 0.63 |
| 752.72 | 1.37 | 0.27 | 2.21 | 0.61 |
| 754.31 | 1.37 | 0.27 | 2.20 | 0.59 |
| 755.90 | 1.37 | 0.27 | 2.19 | 0.58 |
| 757.49 | 1.37 | 0.27 | 2.18 | 0.57 |
| 759.08 | 1.37 | 0.28 | 2.17 | 0.55 |
| 760.67 | 1.37 | 0.28 | 2.16 | 0.54 |
| 762.26 | 1.37 | 0.28 | 2.15 | 0.54 |
| 763.85 | 1.37 | 0.29 | 2.14 | 0.53 |
| 765.44 | 1.37 | 0.29 | 2.13 | 0.52 |
| 767.03 | 1.37 | 0.29 | 2.12 | 0.51 |
| 768.62 | 1.37 | 0.30 | 2.11 | 0.51 |
| 770.21 | 1.37 | 0.30 | 2.10 | 0.51 |
| 771.80 | 1.37 | 0.30 | 2.09 | 0.50 |
| 773.39 | 1.37 | 0.30 | 2.08 | 0.50 |
| 774.98 | 1.37 | 0.31 | 2.07 | 0.50 |
| 776.57 | 1.37 | 0.31 | 2.06 | 0.50 |
| 778.16 | 1.37 | 0.31 | 2.05 | 0.50 |
| 779.75 | 1.38 | 0.32 | 2.04 | 0.50 |
| 781.33 | 1.38 | 0.32 | 2.03 | 0.50 |
| 782.92 | 1.38 | 0.32 | 2.02 | 0.51 |
| 784.51 | 1.38 | 0.32 | 2.01 | 0.51 |
| 786.10 | 1.38 | 0.33 | 2.00 | 0.51 |
| 787.69 | 1.38 | 0.33 | 2.00 | 0.52 |
| 789.28 | 1.38 | 0.33 | 1.99 | 0.52 |
| 790.87 | 1.38 | 0.33 | 1.98 | 0.53 |
| 792.46 | 1.38 | 0.34 | 1.98 | 0.54 |
| 794.04 | 1.39 | 0.34 | 1.97 | 0.55 |
| 795.63 | 1.39 | 0.34 | 1.97 | 0.55 |
| 797.22 | 1.39 | 0.34 | 1.97 | 0.56 |
| 798.81 | 1.39 | 0.34 | 1.97 | 0.57 |
| 800.40 | 1.39 | 0.35 | 1.97 | 0.58 |
| 801.98 | 1.39 | 0.35 | 1.97 | 0.59 |
| 803.57 | 1.39 | 0.35 | 1.97 | 0.59 |
| 805.16 | 1.39 | 0.35 | 1.98 | 0.60 |
| 806.75 | 1.40 | 0.35 | 1.99 | 0.60 |
| 808.33 | 1.40 | 0.36 | 1.99 | 0.61 |
| 809.92 | 1.40 | 0.36 | 2.00 | 0.61 |
| 811.51 | 1.40 | 0.36 | 2.01 | 0.61 |
| 813.09 | 1.40 | 0.36 | 2.01 | 0.60 |
| 814.68 | 1.40 | 0.36 | 2.02 | 0.60 |
| 816.27 | 1.40 | 0.37 | 2.02 | 0.59 |
| 817.86 | 1.40 | 0.37 | 2.02 | 0.59 |
| 819.44 | 1.41 | 0.37 | 2.02 | 0.58 |
| 821.03 | 1.41 | 0.37 | 2.02 | 0.58 |
| 822.61 | 1.41 | 0.37 | 2.02 | 0.57 |
| 824.20 | 1.41 | 0.37 | 2.02 | 0.57 |
| 825.79 | 1.41 | 0.37 | 2.01 | 0.56 |
| 827.37 | 1.41 | 0.38 | 2.01 | 0.56 |
| 828.96 | 1.41 | 0.38 | 2.01 | 0.56 |
| 830.54 | 1.41 | 0.38 | 2.00 | 0.55 |
| 832.13 | 1.42 | 0.38 | 2.00 | 0.55 |
| 833.72 | 1.42 | 0.38 | 1.99 | 0.55 |
| 835.30 | 1.42 | 0.38 | 1.99 | 0.55 |
| 836.89 | 1.42 | 0.38 | 1.98 | 0.55 |
| 838.47 | 1.42 | 0.39 | 1.98 | 0.55 |
| 840.06 | 1.42 | 0.39 | 1.97 | 0.55 |
| 841.64 | 1.42 | 0.39 | 1.97 | 0.55 |
| 843.23 | 1.42 | 0.39 | 1.96 | 0.55 |
| 844.81 | 1.42 | 0.39 | 1.96 | 0.55 |
| 846.39 | 1.42 | 0.39 | 1.95 | 0.55 |
| 847.98 | 1.42 | 0.39 | 1.95 | 0.56 |
| 849.56 | 1.43 | 0.40 | 1.95 | 0.56 |
| 851.15 | 1.43 | 0.40 | 1.94 | 0.56 |
| 852.73 | 1.43 | 0.40 | 1.94 | 0.56 |
| 854.31 | 1.43 | 0.40 | 1.93 | 0.56 |
| 855.90 | 1.43 | 0.40 | 1.93 | 0.57 |
| 857.48 | 1.43 | 0.40 | 1.93 | 0.57 |
| 859.07 | 1.43 | 0.40 | 1.92 | 0.57 |
| 860.65 | 1.43 | 0.41 | 1.92 | 0.57 |
| 862.23 | 1.43 | 0.41 | 1.92 | 0.58 |
| 863.81 | 1.43 | 0.41 | 1.92 | 0.58 |
| 865.40 | 1.43 | 0.41 | 1.91 | 0.58 |
| 866.98 | 1.43 | 0.41 | 1.91 | 0.58 |
| 868.56 | 1.43 | 0.41 | 1.91 | 0.59 |
| 870.14 | 1.43 | 0.41 | 1.90 | 0.59 |
| 871.73 | 1.43 | 0.42 | 1.90 | 0.59 |
| 873.31 | 1.43 | 0.42 | 1.90 | 0.60 |
| 874.89 | 1.43 | 0.42 | 1.90 | 0.60 |
| 876.47 | 1.44 | 0.42 | 1.89 | 0.60 |
| 878.05 | 1.44 | 0.42 | 1.89 | 0.61 |
| 879.64 | 1.44 | 0.42 | 1.89 | 0.61 |
| 881.22 | 1.44 | 0.43 | 1.89 | 0.61 |
| 882.80 | 1.44 | 0.43 | 1.89 | 0.61 |
| 884.38 | 1.44 | 0.43 | 1.88 | 0.62 |
| 885.96 | 1.44 | 0.43 | 1.88 | 0.62 |
| 887.54 | 1.44 | 0.43 | 1.88 | 0.62 |
| 889.12 | 1.44 | 0.43 | 1.88 | 0.63 |
| 890.70 | 1.44 | 0.44 | 1.87 | 0.63 |
| 892.28 | 1.44 | 0.44 | 1.87 | 0.63 |
| 893.86 | 1.44 | 0.44 | 1.87 | 0.64 |
| 895.44 | 1.44 | 0.44 | 1.87 | 0.64 |
| 897.02 | 1.44 | 0.45 | 1.87 | 0.64 |
| 898.60 | 1.44 | 0.45 | 1.86 | 0.65 |
| 900.18 | 1.44 | 0.45 | 1.86 | 0.65 |
| 901.76 | 1.44 | 0.45 | 1.86 | 0.65 |
| 903.34 | 1.44 | 0.45 | 1.86 | 0.66 |
| 904.92 | 1.45 | 0.46 | 1.86 | 0.66 |
| 906.50 | 1.45 | 0.46 | 1.86 | 0.67 |
| 908.08 | 1.45 | 0.46 | 1.85 | 0.67 |
| 909.65 | 1.45 | 0.46 | 1.85 | 0.67 |
| 911.23 | 1.45 | 0.47 | 1.85 | 0.68 |
| 912.81 | 1.45 | 0.47 | 1.85 | 0.68 |
| 914.39 | 1.45 | 0.47 | 1.85 | 0.68 |
| 915.97 | 1.45 | 0.47 | 1.84 | 0.69 |
| 917.54 | 1.45 | 0.48 | 1.84 | 0.69 |
| 919.12 | 1.45 | 0.48 | 1.84 | 0.70 |
| 920.70 | 1.45 | 0.48 | 1.84 | 0.70 |
| 922.28 | 1.46 | 0.48 | 1.84 | 0.71 |
| 923.85 | 1.46 | 0.49 | 1.84 | 0.71 |
| 925.43 | 1.46 | 0.49 | 1.84 | 0.72 |
| 927.01 | 1.46 | 0.49 | 1.83 | 0.72 |
| 928.58 | 1.46 | 0.50 | 1.83 | 0.72 |
| 930.16 | 1.46 | 0.50 | 1.83 | 0.73 |
| 931.73 | 1.47 | 0.50 | 1.83 | 0.73 |
| 933.31 | 1.47 | 0.50 | 1.83 | 0.74 |
| 934.89 | 1.47 | 0.51 | 1.83 | 0.74 |
| 936.46 | 1.47 | 0.51 | 1.83 | 0.75 |
| 938.04 | 1.47 | 0.51 | 1.82 | 0.75 |
| 939.61 | 1.47 | 0.52 | 1.82 | 0.76 |
| 941.19 | 1.48 | 0.52 | 1.82 | 0.76 |
| 942.76 | 1.48 | 0.52 | 1.82 | 0.77 |
| 944.33 | 1.48 | 0.52 | 1.82 | 0.78 |
| 945.91 | 1.48 | 0.53 | 1.82 | 0.78 |
| 947.48 | 1.49 | 0.53 | 1.82 | 0.79 |
| 949.06 | 1.49 | 0.53 | 1.82 | 0.79 |
| 950.63 | 1.49 | 0.54 | 1.82 | 0.80 |
| 952.20 | 1.49 | 0.54 | 1.82 | 0.80 |
| 953.78 | 1.50 | 0.54 | 1.82 | 0.81 |
| 955.35 | 1.50 | 0.54 | 1.82 | 0.82 |
| 956.92 | 1.50 | 0.55 | 1.82 | 0.82 |
| 958.50 | 1.51 | 0.55 | 1.82 | 0.83 |
| 960.07 | 1.51 | 0.55 | 1.82 | 0.83 |
| 961.64 | 1.51 | 0.55 | 1.82 | 0.84 |
| 963.21 | 1.52 | 0.56 | 1.82 | 0.84 |
| 964.79 | 1.52 | 0.56 | 1.82 | 0.85 |
| 966.36 | 1.52 | 0.56 | 1.82 | 0.86 |
| 967.93 | 1.53 | 0.56 | 1.82 | 0.86 |
| 969.50 | 1.53 | 0.57 | 1.82 | 0.87 |
| 971.07 | 1.53 | 0.57 | 1.82 | 0.87 |
| 972.64 | 1.54 | 0.57 | 1.83 | 0.88 |
| 974.21 | 1.54 | 0.57 | 1.83 | 0.89 |
| 975.78 | 1.54 | 0.57 | 1.83 | 0.89 |
| 977.35 | 1.55 | 0.58 | 1.83 | 0.90 |
| 978.92 | 1.55 | 0.58 | 1.83 | 0.90 |
| 980.49 | 1.56 | 0.58 | 1.83 | 0.91 |
| 982.06 | 1.56 | 0.58 | 1.83 | 0.92 |
| 983.63 | 1.56 | 0.58 | 1.84 | 0.92 |
| 985.20 | 1.57 | 0.58 | 1.84 | 0.93 |
| 986.77 | 1.57 | 0.58 | 1.84 | 0.93 |
| 988.34 | 1.58 | 0.59 | 1.84 | 0.94 |
| 989.91 | 1.58 | 0.59 | 1.84 | 0.94 |
| 991.48 | 1.59 | 0.59 | 1.84 | 0.95 |
| 993.04 | 1.59 | 0.59 | 1.85 | 0.96 |
| 994.61 | 1.60 | 0.59 | 1.85 | 0.96 |
| 996.18 | 1.60 | 0.59 | 1.85 | 0.97 |
| 997.75 | 1.60 | 0.59 | 1.85 | 0.97 |
| 999.31 | 1.61 | 0.59 | 1.85 | 0.98 |
| 1013.13 | 1.65 | 0.59 | 1.87 | 1.04 |
| 1016.54 | 1.66 | 0.59 | 1.87 | 1.06 |
| 1019.94 | 1.67 | 0.59 | 1.88 | 1.08 |
| 1023.35 | 1.68 | 0.58 | 1.89 | 1.10 |
| 1026.75 | 1.69 | 0.58 | 1.90 | 1.12 |
| 1030.16 | 1.70 | 0.58 | 1.91 | 1.14 |
| 1033.57 | 1.71 | 0.57 | 1.93 | 1.17 |
| 1036.97 | 1.71 | 0.57 | 1.96 | 1.19 |
| 1040.38 | 1.72 | 0.56 | 1.99 | 1.20 |
| 1043.79 | 1.73 | 0.56 | 2.03 | 1.21 |
| 1047.20 | 1.74 | 0.55 | 2.07 | 1.21 |
| 1050.61 | 1.74 | 0.55 | 2.10 | 1.20 |
| 1054.02 | 1.75 | 0.54 | 2.13 | 1.18 |
| 1057.43 | 1.76 | 0.53 | 2.15 | 1.15 |
| 1060.84 | 1.76 | 0.53 | 2.15 | 1.12 |
| 1064.25 | 1.77 | 0.52 | 2.15 | 1.09 |
| 1067.66 | 1.77 | 0.51 | 2.14 | 1.07 |
| 1071.07 | 1.78 | 0.51 | 2.12 | 1.06 |
| 1074.48 | 1.78 | 0.50 | 2.10 | 1.05 |
| 1077.89 | 1.79 | 0.49 | 2.07 | 1.05 |
| 1081.30 | 1.79 | 0.49 | 2.04 | 1.06 |
| 1084.71 | 1.79 | 0.48 | 2.02 | 1.08 |
| 1088.13 | 1.79 | 0.47 | 1.99 | 1.10 |
| 1091.54 | 1.80 | 0.47 | 1.97 | 1.13 |
| 1094.95 | 1.80 | 0.46 | 1.95 | 1.16 |
| 1098.37 | 1.80 | 0.45 | 1.94 | 1.21 |
| 1101.78 | 1.80 | 0.45 | 1.92 | 1.26 |
| 1105.20 | 1.80 | 0.44 | 1.92 | 1.31 |
| 1108.61 | 1.80 | 0.43 | 1.93 | 1.37 |
| 1112.03 | 1.80 | 0.43 | 1.94 | 1.43 |
| 1115.44 | 1.80 | 0.42 | 1.97 | 1.50 |
| 1118.86 | 1.80 | 0.41 | 2.01 | 1.57 |
| 1122.27 | 1.80 | 0.41 | 2.06 | 1.63 |
| 1125.69 | 1.80 | 0.40 | 2.12 | 1.69 |
| 1129.11 | 1.80 | 0.40 | 2.20 | 1.74 |
| 1132.52 | 1.80 | 0.39 | 2.29 | 1.78 |
| 1135.94 | 1.80 | 0.39 | 2.39 | 1.80 |
| 1139.36 | 1.80 | 0.38 | 2.49 | 1.80 |
| 1142.78 | 1.80 | 0.37 | 2.59 | 1.78 |
| 1146.20 | 1.80 | 0.37 | 2.68 | 1.75 |
| 1149.61 | 1.80 | 0.36 | 2.76 | 1.70 |
| 1153.03 | 1.80 | 0.36 | 2.83 | 1.63 |
| 1156.45 | 1.79 | 0.36 | 2.87 | 1.57 |
| 1159.87 | 1.79 | 0.35 | 2.91 | 1.50 |
| 1163.29 | 1.79 | 0.35 | 2.92 | 1.43 |
| 1166.71 | 1.79 | 0.34 | 2.93 | 1.38 |
| 1170.14 | 1.79 | 0.34 | 2.92 | 1.33 |
| 1173.56 | 1.78 | 0.33 | 2.91 | 1.29 |
| 1176.98 | 1.78 | 0.33 | 2.90 | 1.27 |
| 1180.40 | 1.78 | 0.33 | 2.89 | 1.26 |
| 1183.82 | 1.78 | 0.32 | 2.89 | 1.26 |
| 1187.25 | 1.77 | 0.32 | 2.90 | 1.27 |
| 1190.67 | 1.77 | 0.32 | 2.92 | 1.27 |
| 1194.09 | 1.77 | 0.31 | 2.96 | 1.27 |
| 1197.52 | 1.77 | 0.31 | 3.01 | 1.25 |
| 1200.94 | 1.76 | 0.31 | 3.05 | 1.22 |
| 1204.36 | 1.76 | 0.31 | 3.09 | 1.17 |
| 1207.79 | 1.76 | 0.30 | 3.11 | 1.11 |
| 1211.21 | 1.75 | 0.30 | 3.12 | 1.04 |
| 1214.64 | 1.75 | 0.30 | 3.12 | 0.98 |
| 1218.07 | 1.75 | 0.30 | 3.10 | 0.93 |
| 1221.49 | 1.75 | 0.29 | 3.07 | 0.89 |
| 1224.92 | 1.74 | 0.29 | 3.04 | 0.85 |
| 1228.35 | 1.74 | 0.29 | 3.01 | 0.82 |
| 1231.77 | 1.74 | 0.29 | 2.97 | 0.81 |
| 1235.20 | 1.73 | 0.29 | 2.94 | 0.80 |
| 1238.63 | 1.73 | 0.29 | 2.91 | 0.79 |
| 1242.06 | 1.73 | 0.28 | 2.88 | 0.79 |
| 1245.49 | 1.72 | 0.28 | 2.85 | 0.80 |
| 1248.91 | 1.72 | 0.28 | 2.83 | 0.82 |
| 1252.34 | 1.72 | 0.28 | 2.81 | 0.84 |
| 1255.77 | 1.72 | 0.28 | 2.80 | 0.86 |
| 1259.20 | 1.71 | 0.28 | 2.79 | 0.88 |
| 1262.63 | 1.71 | 0.28 | 2.79 | 0.91 |
| 1266.06 | 1.71 | 0.28 | 2.80 | 0.94 |
| 1269.50 | 1.70 | 0.27 | 2.81 | 0.97 |
| 1272.93 | 1.70 | 0.27 | 2.84 | 1.00 |
| 1276.36 | 1.70 | 0.27 | 2.87 | 1.02 |
| 1279.79 | 1.69 | 0.27 | 2.91 | 1.04 |

| 1283.22 | 1.69 | 0.27 | 2.95 | 1.05 |
|---------|------|------|------|------|
| 1286.66 | 1.69 | 0.27 | 3.00 | 1.05 |
| 1290.09 | 1.68 | 0.27 | 3.04 | 1.04 |
| 1293.52 | 1.68 | 0.27 | 3.09 | 1.02 |
| 1296.96 | 1.68 | 0.27 | 3.13 | 1.00 |
| 1300.39 | 1.68 | 0.27 | 3.17 | 0.96 |
| 1303.83 | 1.67 | 0.27 | 3.21 | 0.93 |
| 1307.26 | 1.67 | 0.27 | 3.23 | 0.88 |
| 1310.70 | 1.67 | 0.27 | 3.25 | 0.84 |
| 1314.13 | 1.66 | 0.27 | 3.27 | 0.80 |
| 1317.57 | 1.66 | 0.27 | 3.28 | 0.75 |
| 1321.00 | 1.66 | 0.27 | 3.28 | 0.71 |
| 1324.44 | 1.65 | 0.27 | 3.28 | 0.67 |
| 1327.88 | 1.65 | 0.27 | 3.27 | 0.63 |
| 1331.32 | 1.65 | 0.27 | 3.27 | 0.60 |
| 1334.75 | 1.65 | 0.27 | 3.26 | 0.56 |
| 1338.19 | 1.64 | 0.27 | 3.25 | 0.53 |
| 1341.63 | 1.64 | 0.27 | 3.23 | 0.50 |
| 1345.07 | 1.64 | 0.27 | 3.22 | 0.48 |
| 1348.51 | 1.63 | 0.27 | 3.21 | 0.45 |
| 1351.95 | 1.63 | 0.27 | 3.19 | 0.43 |
| 1355.39 | 1.63 | 0.28 | 3.18 | 0.41 |
| 1358.83 | 1.63 | 0.28 | 3.16 | 0.39 |
| 1362.27 | 1.62 | 0.28 | 3.15 | 0.38 |
| 1365.71 | 1.62 | 0.28 | 3.13 | 0.36 |
| 1369.15 | 1.62 | 0.28 | 3.12 | 0.35 |
| 1372.59 | 1.61 | 0.28 | 3.10 | 0.33 |
| 1376.03 | 1.61 | 0.28 | 3.09 | 0.32 |
| 1379.48 | 1.61 | 0.28 | 3.08 | 0.31 |
| 1382.92 | 1.61 | 0.28 | 3.06 | 0.30 |
| 1386.36 | 1.60 | 0.29 | 3.05 | 0.29 |
| 1389.81 | 1.60 | 0.29 | 3.04 | 0.28 |
| 1393.25 | 1.60 | 0.29 | 3.03 | 0.27 |
| 1396.69 | 1.59 | 0.29 | 3.01 | 0.26 |
| 1400.14 | 1.59 | 0.29 | 3.00 | 0.26 |
| 1403.58 | 1.59 | 0.29 | 2.99 | 0.25 |
| 1407.03 | 1.59 | 0.29 | 2.98 | 0.24 |
| 1410.47 | 1.58 | 0.30 | 2.97 | 0.24 |
| 1413.92 | 1.58 | 0.30 | 2.96 | 0.23 |
| 1417.36 | 1.58 | 0.30 | 2.95 | 0.22 |
| 1420.81 | 1.58 | 0.30 | 2.94 | 0.22 |
| 1424.26 | 1.57 | 0.30 | 2.93 | 0.21 |
| 1427.71 | 1.57 | 0.31 | 2.92 | 0.21 |
| 1431.15 | 1.57 | 0.31 | 2.91 | 0.21 |
| 1434.60 | 1.57 | 0.31 | 2.90 | 0.20 |
| 1438.05 | 1.56 | 0.31 | 2.89 | 0.20 |
| 1441.50 | 1.56 | 0.31 | 2.88 | 0.19 |
| 1444.95 | 1.56 | 0.32 | 2.87 | 0.19 |
| 1448.40 | 1.56 | 0.32 | 2.87 | 0.19 |
| 1451.85 | 1.55 | 0.32 | 2.86 | 0.18 |
| 1455.30 | 1.55 | 0.32 | 2.85 | 0.18 |
| 1458.75 | 1.55 | 0.33 | 2.84 | 0.18 |
| 1462.20 | 1.55 | 0.33 | 2.84 | 0.17 |
| 1465.65 | 1.55 | 0.33 | 2.83 | 0.17 |
| 1469.10 | 1.54 | 0.33 | 2.82 | 0.17 |
| 1472.55 | 1.54 | 0.34 | 2.81 | 0.17 |
| 1476.00 | 1.54 | 0.34 | 2.81 | 0.16 |
| 1479.45 | 1.54 | 0.34 | 2.80 | 0.16 |
| 1482.91 | 1.54 | 0.35 | 2.79 | 0.16 |
| 1486.36 | 1.53 | 0.35 | 2.79 | 0.16 |
| 1489.81 | 1.53 | 0.35 | 2.78 | 0.15 |
| 1493.27 | 1.53 | 0.35 | 2.77 | 0.15 |
| 1496.72 | 1.53 | 0.36 | 2.77 | 0.15 |
| 1500.18 | 1.53 | 0.36 | 2.76 | 0.15 |
| 1503.63 | 1.53 | 0.36 | 2.76 | 0.15 |
| 1507.09 | 1.52 | 0.37 | 2.75 | 0.14 |
| 1510.54 | 1.52 | 0.37 | 2.75 | 0.14 |
| 1514.00 | 1.52 | 0.37 | 2.74 | 0.14 |
| 1517.45 | 1.52 | 0.38 | 2.73 | 0.14 |
| 1520.91 | 1.52 | 0.38 | 2.73 | 0.14 |
| 1524.37 | 1.52 | 0.38 | 2.72 | 0.14 |
| 1527.83 | 1.52 | 0.39 | 2.72 | 0.13 |
| 1531.28 | 1.51 | 0.39 | 2.71 | 0.13 |
| 1534.74 | 1.51 | 0.39 | 2.71 | 0.13 |
| 1538.20 | 1.51 | 0.40 | 2.70 | 0.13 |
| 1541.66 | 1.51 | 0.40 | 2.70 | 0.13 |
| 1545.12 | 1.51 | 0.41 | 2.69 | 0.13 |
| 1548.58 | 1.51 | 0.41 | 2.69 | 0.13 |
| 1552.04 | 1.51 | 0.41 | 2.68 | 0.12 |
| 1555.50 | 1.51 | 0.42 | 2.68 | 0.12 |
| 1558.96 | 1.51 | 0.42 | 2.67 | 0.12 |
| 1562.42 | 1.51 | 0.43 | 2.67 | 0.12 |
| 1565.88 | 1.51 | 0.43 | 2.67 | 0.12 |
| 1569.34 | 1.50 | 0.43 | 2.66 | 0.12 |
| 1572.80 | 1.50 | 0.44 | 2.66 | 0.12 |
| 1576.26 | 1.50 | 0.44 | 2.65 | 0.12 |
| 1579.73 | 1.50 | 0.44 | 2.65 | 0.12 |
| 1583.19 | 1.50 | 0.45 | 2.64 | 0.11 |
| 1586.65 | 1.50 | 0.45 | 2.64 | 0.11 |
| 1590.12 | 1.50 | 0.46 | 2.64 | 0.11 |
| 1593.58 | 1.50 | 0.46 | 2.63 | 0.11 |
| 1597.04 | 1.50 | 0.47 | 2.63 | 0.11 |
| 1600.51 | 1.50 | 0.47 | 2.62 | 0.11 |
| 1603.97 | 1.50 | 0.47 | 2.62 | 0.11 |
| 1607.44 | 1.50 | 0.48 | 2.62 | 0.11 |
| 1610.91 | 1.50 | 0.48 | 2.61 | 0.11 |
| 1614.37 | 1.51 | 0.49 | 2.61 | 0.11 |
| 1617.84 | 1.51 | 0.49 | 2.60 | 0.11 |
| 1621.30 | 1.51 | 0.49 | 2.60 | 0.11 |
| 1624.77 | 1.51 | 0.50 | 2.60 | 0.10 |
| 1628.24 | 1.51 | 0.50 | 2.59 | 0.10 |
| 1631.71 | 1.51 | 0.51 | 2.59 | 0.10 |
| 1635.18 | 1.51 | 0.51 | 2.59 | 0.10 |
| 1638.64 | 1.51 | 0.52 | 2.58 | 0.10 |
| 1642.11 | 1.51 | 0.52 | 2.58 | 0.10 |
| 1645.58 | 1.51 | 0.52 | 2.58 | 0.10 |
| 1649.05 | 1.52 | 0.53 | 2.57 | 0.10 |
| 1652.52 | 1.52 | 0.53 | 2.57 | 0.10 |
| 1655.99 | 1.52 | 0.54 | 2.57 | 0.10 |
| 1659.46 | 1.52 | 0.54 | 2.56 | 0.10 |
| 1662.93 | 1.52 | 0.54 | 2.56 | 0.10 |
| 1666.40 | 1.52 | 0.55 | 2.56 | 0.10 |
| 1669.88 | 1.53 | 0.55 | 2.55 | 0.10 |
| 1673.35 | 1.53 | 0.56 | 2.55 | 0.09 |
| 1676.82 | 1.53 | 0.56 | 2.55 | 0.09 |
| 1680.29 | 1.53 | 0.57 | 2.54 | 0.09 |
| 1683.77 | 1.53 | 0.57 | 2.54 | 0.09 |
| 1687.24 | 1.54 | 0.57 | 2.54 | 0.09 |